\pdfoutput=1
\documentclass[a4paper,american, superscriptaddress, amsmath, amssymb,
reprint, aip, jcp
]{revtex4-1}
\usepackage{lmodern}

\usepackage[T1]{fontenc} \usepackage[utf8]{inputenc}
\usepackage{float}
\usepackage{amsmath}
\usepackage{amssymb}
\usepackage{graphicx}
\usepackage{esint}

\makeatletter

\pdfpageheight\paperheight \pdfpagewidth\paperwidth

\floatstyle{ruled}
\newfloat{algorithm}{tbp}{loa}
\providecommand{\algorithmname}{Algorithm}
\floatname{algorithm}{\protect\algorithmname}

\@ifundefined{textcolor}{} {%
 \definecolor{BLACK}{gray}{0}
 \definecolor{WHITE}{gray}{1}
 \definecolor{RED}{rgb}{1,0,0}
 \definecolor{GREEN}{rgb}{0,1,0}
 \definecolor{BLUE}{rgb}{0,0,1}
 \definecolor{CYAN}{cmyk}{1,0,0,0}
 \definecolor{MAGENTA}{cmyk}{0,1,0,0}
 \definecolor{YELLOW}{cmyk}{0,0,1,0}
}

\makeatother

\usepackage{babel}
\begin{document}
\global\long\def\ts#1{_{\text{#1}}} \global\long\def\tsup#1{^{\text{#1}}}

\global\long\def\E#1{\langle#1\rangle}
\global\long\def\bE#1{\bigl\langle#1\bigr\rangle}
\global\long\def\BE#1{\Bigl\langle#1\Bigr\rangle}
\global\long\def\dd{\mathrm{d}} \global\long\def\roi{\mathrm{\mathcal{R}}}

\title{Non-Stationary Forward Flux Sampling}

\author{Nils B. Becker}

\affiliation{FOM Institute for Atomic and Molecular Physics (AMOLF), Science
Park 104, 1098 XG Amsterdam, The Netherlands}

\author{Rosalind J. Allen}

\affiliation{SUPA, School of Physics and Astronomy, The University of
Edinburgh, James Clerk Maxwell Building, The King's Buildings, Mayfield Road,
Edinburgh EH9 3JZ, UK}

\author{Pieter Rein ten Wolde}

\affiliation{FOM Institute for Atomic and Molecular Physics (AMOLF), Science
Park 104, 1098 XG Amsterdam, The Netherlands}
\begin{abstract}
 We present a new method, Non-Stationary Forward Flux Sampling, that allows
 efficient  simulation of rare events in both stationary and non-stationary
 stochastic systems. The method uses  stochastic branching and pruning
 to achieve uniform sampling of trajectories in phase space and time, leading
 to accurate estimates for time-dependent switching propensities and
 time-dependent phase space probability densities. The method is suitable for
 equilibrium or non-equilibrium systems, in or out of stationary state,
 including non-Markovian or externally driven systems. We demonstrate the
 validity of the technique by applying it to a one-dimensional barrier crossing
 problem that can be solved exactly, and show its usefulness by applying it to
 the time-dependent switching of a genetic toggle switch.
\end{abstract}
\maketitle

\section{Introduction}

Rare events, which occur infrequently but have important consequences, control
the dynamical behavior of many physical systems, both in and out of
equilibrium -- classic examples include crystal nucleation, protein folding,
earthquakes and traffic jams. When simulating such systems on a computer, some
form of enhanced sampling is usually needed in order to generate any
significant number of rare event samples on the time scale of the simulation.
While a number of enhanced sampling methods are available for systems in
steady state, many important rare event processes happen in non-stationary
systems, for which most existing methods are unsuitable.
In this article, we introduce a new enhanced sampling
scheme, Non-Stationary Forward Flux Sampling,  which allows efficient
simulation of rare events in  both stationary and non-stationary stochastic
systems.

For systems in thermodynamic equilibrium, a large variety of rare-event
techniques have been developed. One is the Bennett-Chandler method
\cite{bennett77,chandler78}, which involves a calculation of the free energy
along a predetermined reaction coordinate, followed by a computation of the
kinetic pre-factor by firing off trajectories from the top of the free-energy
barrier. Other techniques are transition path sampling \cite{bolhuis02} and
transition interface sampling (TIS) \cite{erp03}, which employ Monte Carlo
sampling of the ensemble of transition paths, approximate schemes such as
partial-path TIS \cite{moroni04} and milestoning \cite{faradjian04}, which use
a series of interfaces in phase space between the initial and final state, and
string methods \cite{e02}. While these schemes have been successfully applied
to a large class of problems, they do require knowledge of the phase-space
density, which  limits their use to systems in thermodynamic equilibrium.

For non-equilibrium systems, the phase-space density is generally not known. 
This severely limits the possibilities for
devising enhanced sampling schemes to calculate transition rates. Yet, for
non-equilibrium systems that are in stationary state, recently a number of
rare-event techniques have been developed. One is the weighted-ensemble (WE)
method \cite{huber96,zhang10a}, where phase space is divided into bins, and
trajectories are selected and re-weighted bin-wise to achieve uniform coverage
of the phase space. Another technique is the non-equilibrium minimum action
method \cite{heymann08}, which allows the characterization of transition paths
but not rate constants. Non-equilibrium umbrella sampling \cite{warmflash07}
coarse-grains systems with Markovian dynamics on overlapping grids in state
space and biases inter- vs.~intra-bin transitions. Forward-Flux Sampling (FFS)
\cite{allen06a} uses a series of interfaces in phase space between the initial
and final state to drive the system over the barrier in a ratchet-like manner,
by capitalizing on those fluctuations that move the system from one interface
to the next. While these methods do not require thermal equilibrium, they
rely on the system being in stationary state.

In reality, however, many important rare event processes happen in systems
which are not in stationary state. For these processes, the propensity
(probability per unit time) for the rare event to occur is time-dependent; this
time dependence may be caused by external driving, by transient relaxation of
the system from an out-of-equilibrium initial state, or by the presence of
memory in the dynamics on a relevant time scale.
In fact many real-life instances of the rare event processes mentioned above
are time-dependent, such as: crystal nucleation during flash-freezing
(e.g.~when preparing cryo-electron microscopy samples); protein folding during
transient association with a chaperone protein; and triggering of traffic jams
by brief disturbances on the road. Other interesting cases  include transitions
between multiple limit cycles in neural networks under time-dependent stimuli
(as suggested for epileptic seizures, e.g.~\cite{osorio11}), and the response
of metastable biochemical networks to transient signals, e.g.~in cell
differentiation \cite{sueel06} or in viral life cycle progression, see
sec.~\ref{sec:Applications}.
These important types of rare events are not accessible to any existing
enhanced-sampling techniques, with the exception of the FFS-inspired method of
Berryman and Schilling \cite{berryman10a} which relies on mapping the systems
dynamics onto a time-inhomogeneous Markov process. The noise-sampling method of
Crooks and Chandler \cite{crooks01} allows sampling of transition paths in
non-stationary systems but cannot be used to compute time-dependent transition
rates, since that would require knowledge of the initial phase-space density,
which is unavailable in general.

In this article, we describe a new method, Non-Stationary Forward Flux Sampling
(NS-FFS), that allows efficient simulation of rare events in time-dependent
stochastic dynamical systems.  NS-FFS  constitutes a  time-dependent
generalization of FFS, and is conceptually straightforward and
easy-to-implement.
NS-FFS achieves uniform sampling of trajectories crossing a predefined region
in time and phase space, by combining interfaces in phase space as used in FFS
\cite{allen06a, allen09} with a flat-histogram pruned-enriched Rosenbluth
method originally developed for polymer simulations
\cite{grassberger97,prellberg04}. In NS-FFS, trajectories are branched
(proliferated) or pruned (terminated) based on their progression towards the
final state, using interfaces in phase space \emph{and} time.  The scheme can
be employed to sample the time-dependent phase-space density and time-dependent
crossing fluxes, with uniform relative error.  It thereby gives access to
time-dependent transition rate functions \cite{becker12}, including their
low-propensity tails.

The article is structured as follows. In section
\ref{sec:Stationary-vs.non-stationary} we provide a theoretical background,
contrasting the well-studied setting of stationary, Markovian barrier-crossing
with more general time-dependent rare event problems.  Section
\ref{sec:Nonstationary-Flux-Sampling} presents the NS-FFS algorithm, together
with corresponding pseudo-code. The correctness and efficiency of the algorithm
are demonstrated in Section \ref{sec:Applications} using two simple examples:
diffusive escape in a one-dimensional W-shaped potential, and time-dependent
switching in a genetic toggle switch.  We conclude by discussing the main
features of the method, and possible extensions, in Section
\ref{sec:Discussion}.

\section{Time-dependent rare events}\label{sec:Stationary-vs.non-stationary}

% \subsection{Rate constants for Markovian transitions}
\subsection{Transition rate constants for Markovian 
systems in stationary state}

\begin{figure}
\begin{centering}
\includegraphics[width=6cm]{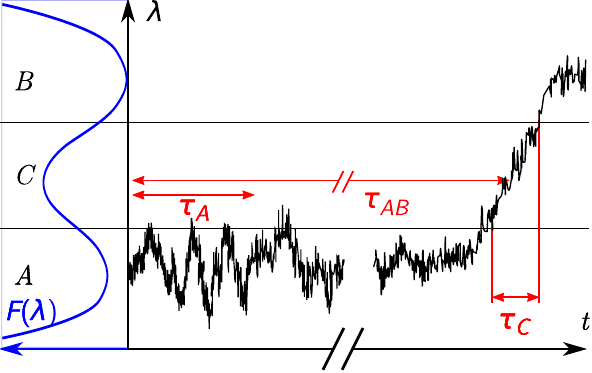}
\par\end{centering}

\caption{\label{fig:timescales} Barrier crossing in stationary state. 
Equilibration within the metastable region $A$ occurs within $\tau_{A}$. After
a mean waiting time $\tau_{AB}$ transitions from $A$ to $B$ occur,
with a typical duration $\tau_{C}$.}
\end{figure}

We first discuss rare transitions occurring
in a system in stationary state. They may be visualized in terms of a 
static free-energy landscape 
\footnote{The term `free energy' is used in the
  sense of a negative logarithm of the density; this remains meaningful also
  for non-equilibrium steady states.%
}, as shown in Fig.~\ref{fig:timescales}. 
A typical trajectory starts inside the metastable state $A$ and fully
explores the basin within a time $\tau_{A}$; after a waiting time $\tau_{AB}$
it makes a rapid transition (on a timescale $\tau_{C}$) over a high free-energy
barrier $C$ into state $B$. 
The essential observation is that if the
equilibration time $\tau_{A}$ is much shorter than the mean waiting time in the
$A$ basin $\tau_{AB}$, then in the regime $\tau_{A},\tau_{C}\ll t\ll\tau_{AB},$
transitions from $A$ to $B$ occur in a Markovian, memoryless fashion,
effectively starting from a stationary state within $A$. 
Since the system is still in $A$ with probability
$\simeq1$, switches also happen with a constant propensity, whose
value equals the rate constant $k_{AB}=\tau_{AB}^{-1}$ (see also
\cite{becker12}). Numerical techniques  for
simulating rare events in stationary systems exploit this fact by
generating biased ensembles of \emph{short} transition paths of
duration $\tau_A,\tau_C$, which nevertheless allow reliable estimates
of the much \emph{longer} waiting time $\tau_{AB}\gg \tau_A,\tau_C$.

We briefly review how this works in FFS \cite{allen06a}. Given a progress
coordinate $\lambda$ which increases from $A$ to $B$, one defines a set of
interfaces at successive levels $\{\lambda_{l}\}_{1\leq l\leq L}$. A sample of
points at the  initial interface $\lambda_{1}$ is generated by a
quasi-stationary simulation within state $A$.
% note the points at interface 1 are not sampled from the quasi-stattionary
% distribution because they represent the forward flux only not the backward
% flux
These are then used to initialize a set of trajectories
which are propagated to interface $\lambda_{2}$, or terminated if they re-enter
$A$, whichever happens first. This procedure is repeated, starting from
$\lambda_{2}$ and stopping at $\lambda_{3}$ or $A$, and so on. By propagating
trajectories in segments from one $\lambda$-interface to the next, FFS
capitalizes on the rare fluctuations towards the transition. The resulting
transition paths are of length $\tau_{C}\lesssim t\ll\tau_{AB}$. This leads to
efficiency gains which grow exponentially with the barrier height.

\subsection{Time-dependent rare transition events}

In this paper we are interested in
situations where the metastable states
$A$ and $B$ can still be identified, but transitions between them happen with
time-dependent propensity. For example, let us suppose that the generic system
illustrated in Fig.~\ref{fig:timescales} and discussed above is exposed to 
weak external forcing with protocol $\phi(t)$, $0<t<T$. For a macroscopic 
description with a time resolution coarser that $\tau_A$, the system is 
macroscopically Markovian and one can still
define a transition rate from $A$ to $B$, but this now depends on
time: $k_{AB}=k_{AB}(t)$ \cite{becker12}.

Transition events may then be `uniformly rare' so that $k_{AB}^{-1}(t)\gg T$
for all $t$ and the survival probability $S_A(t)\simeq 1$ up to time $T$.
However, if the transition rate is high in a particular time window, then the
survival probability $S_{A}(t)$ may drop significantly below 1 during the time
interval $(0,T)$ of interest (as in figs.~\ref{fig:on-switch},\ref{fig:pulse}
discussed below), and has to be taken into account in computing $k_{AB}(t)$
\cite{becker12}.  One then needs to measure both the first-passage time
distribution or flux $q_{AB}(t)$ from $A$ to $B$ and the survival probability
$S_{A}(t)$, to extract the time-dependent rate $k_{AB}(t)=q_{AB}(t)/S_{A}(t)$;
see the accompanying paper \cite{becker12} for a detailed discussion.

Alternatively, transitions from $A$ to $B$ may be time-dependent even in
systems  without external driving, due to ``macroscopic memory'', in which
 the system's dynamics evolves on a  time scale $\tau\ts{slow}$ such that
$\tau_A<\tau\ts{slow}<\tau_{AB}$. In this case, relaxation within the $A$ basin
will no longer be effectively instantaneous,  and for $t<\tau\ts{slow}$ the
exit propensity from $A$ will depend on the history of the trajectory. While
transitions from $A$ to $B$ cannot be described by a rate constant, one may
be able to characterize such systems in terms of a rate kernel $k_{AB}(t|t')$,
which quantifies the propensity to switch from $A$ to $B$ for the first time at
time $t$, given that the previous switch happened at $t'<t$ (see
\cite{becker12} for a detailed discussion). To extract the rate kernel from a
simulation, one needs to measure the probability to stay in $A$ without
interruption from $t'$ to $t$; this requires a simulation over a time interval
$(0,T)$ where $T\gtrsim\tau\ts{slow}$. In some systems memory effects may be
combined with external driving -- for example barrier escape in an underdamped,
driven system.

In all of these scenarios, the system dynamics is nontrivial and interesting
over a time window $(0,T)$ (determined by the external driving or internal
memory) which is longer than the typical transition time $\tau_{AB}$.   This fact makes
it impossible to speed up the simulation by generating only an ensemble of
short trajectories of length $\gtrsim \tau_A,\tau_C$, as is done in the rare event
techniques for stationary systems discussed above. To capture the physical
behavior of non-stationary systems, trajectories must extend over the
\emph{entire} time window $(0,T)$ of interest.

Nevertheless, enhanced sampling is both useful and feasible for non-stationary
systems. Clearly, if in the time window of interest $0<t<T$ an event occurs
with low probability, then brute-force simulations will fail to generate more
than a few, if any, events on this time scale. The goal of an enhanced sampling
method is therefore to  generate an ensemble of trajectories of full length $T$
which is biased towards the transition. By reweighting this ensemble one can
compute properties such as the time-dependent probability density, the
transition flux $q_{AB}(t)$, the transition rate function $k_{AB}(t)$ or the
transition rate kernel $k_{AB}(t|t')$ for the original system, over the time
window $0<t<T$ of interest. NS-FFS, presented below, is precisely such a
technique: by proliferating trajectories that evolve towards the final state
and terminating those that do not, it generates transition events in the time
window of interest; moreover, the branching/pruning strategy is such that the
relative sampling error is uniform over the space-time region of interest.

\section{Non-stationary Forward Flux Sampling}
\label{sec:Nonstationary-Flux-Sampling}

The aim of the NS-FFS method is to generate a biased set of trajectories
which sample transitions from state $A$ to state $B$, defined by a progress
coordinate $\lambda$, as a function of time, for non-stationary stochastic
systems. To bias the set of trajectories towards the transition, one would like
the flux of trajectories in the biased ensemble to be uniform in $\lambda$ (as
in methods like FFS); to sample accurately the time-dependent behavior of the
system (i.e. to obtain good sampling of early as well as late transition
events), one would also like the flux of trajectories to be uniform in time.
NS-FFS achieves both of these objectives, by generating a set of trajectories
that uniformly cover a specified region of interest $\roi$ in time and in the
progress coordinate. Rare excursions into low-probability regions within $\roi$
are sampled with the same accuracy as common events, and early excursions are
sampled with the same accuracy as late ones.

The method is conceptually simple. First, one
defines the region of $\lambda$-time space $\roi$ that is of interest. One
then partitions the region $\roi$ using a series of interfaces; the interfaces
may be defined as a level set either in the progress coordinate or in the time
coordinate. Each interface is then partitioned into a set of bins; the bins are
defined as intervals in the respective other coordinate (time for $\lambda$
interfaces and $\lambda$ for time interfaces).

The simulations start by generating an ensemble of initial conditions, for
example by performing a brute-force simulation in the initial state $A$.
% TODO
% These serve as an efficient device to both monitor the progress of
% trajectories towards a rare event, and to decompose the extremely small
% probability of
%generating a long trajectory to go from $A$ all the way to the final state $B$,
%into larger probabilities of generating shorter segments of trajectories that
%only go from one interface to the next.
%I think this is distracting here: could maybe be moved somewhere else?
One then proceeds by firing off a
trajectory from a randomly chosen initial condition, and propagating it
according to the given dynamical equations of the system. The trajectory
is assigned a statistical weight, which is initially unity. Upon crossing one
of the pre-defined interfaces, the trajectory may be branched (split into
several `child' trajectories, with new statistical weights), or pruned
(terminated). Repeating this procedure recursively for all child trajectories,
one generates a `trajectory tree' which extends form 0 to $T$. The whole
procedure is then repeated by firing off a new trajectory from a randomly
chosen initial condition.

The probability that a trajectory is branched or pruned when crossing a given
interface bin depends on a running histogram which monitors the weighted flux
of trajectories crossing that bin. The branching/pruning rule is set up such
that trajectories which arrive at a bin which has previously been under-sampled
are likely to branch; those which arrive at a previously over-sampled bin are
likely to be pruned. The algorithm successively approaches a steady state in
which the numbers of trajectories passing through all bins on all interfaces
are equal -- i.e. uniform sampling is achieved both in time and in the progress
coordinate.

In this section, we discuss each of the three ingredients---stochastic
branching and pruning; interfaces in phase-time space; the
flat-histogram rule---in more detail, relegating mathematical details
to Appendices. We then present pseudo-code of the resulting NS-FFS
algorithm. Finally, we comment on the main features of the
method. Again, implementation details are given in the Appendix.

\subsection{Stochastic branching and pruning}

A key element of the NS-FFS algorithm is the branching and pruning of
trajectories, which allows control of the trajectory density. In a branching
move, independent copies of the trajectory are created with a common history up
to time $t$, while in a pruning move, the trajectory is terminated.
The essential observation here is that one is free to branch or prune
trajectories at will, as long as the statistical weights of the child
trajectories are adjusted (reweighted) appropriately.

\begin{figure}
\begin{centering}
\includegraphics[width=8cm]{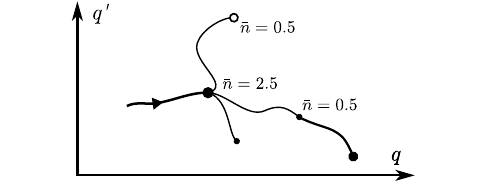}
\par\end{centering}

\caption{\label{fig:branch_reweight}Stochastic branching/pruning. A
branching\slash{}pruning move with average child number $\bar{n}$ proliferates
or terminates branches. If $\bar{n}\gtrless1$, surviving branches are decreased
or increased in weight, respectively (in the diagram, line thickness
represents statistical weight). The target child number $\bar{n}$ may be an
arbitrary function of the  chosen coordinates $q,q'$, as long as the
reweighting factor $r$ is satisfies Eq.~\ref{eq:unbiased}.}
\end{figure}

Suppose that at time $t$, a trajectory
with statistical weight $w$ is randomly branched into $n=1\dots
n\ts{max}$ children with probability $b(n)$, or pruned with
probability $b(n=0)$ (Fig.~\ref{fig:branch_reweight}). 
Each child branch is assigned a new weight $w'=r(n)\times w$. 
Clearly this branching/pruning move will be statistically 
unbiased only if the weight factor $r(n)$ is chosen correctly.

A necessary and sufficient condition for the combination of $b$ and
$r$ to be correct is that weight be conserved on average over
branching/pruning outcomes. That is, we may choose the branch
number distribution and reweighting factor at will as long as they
satisfy
\begin{equation}
\E{nr}=\sum_{n=0}^{n\ts{max}}b(n)nr(n)=1
\text{, where }\sum_{n=0}^{n\ts{max}}b(n)=1.\label{eq:unbiased}
\end{equation}
This holds under very general conditions, including non-stationary
system dynamics with memory and dependence of $b$ on arbitrary parameters,
as shown in app.~\ref{sec:correctness}.

Thus we are free to adjust $b$ to yield a desired mean branch number $\E
n\in(0,n\ts{max})$, and thereby enrich or dilute the density of sample paths
based on any chosen criterion, as long as we also adapt $r$ to satisfy
Eq.~\ref{eq:unbiased}.

\subsection{Interfaces in phase-time space}\label{sub:Interfaces-in-phase-time}

\begin{figure*}
\begin{centering}
\includegraphics[width=13cm]{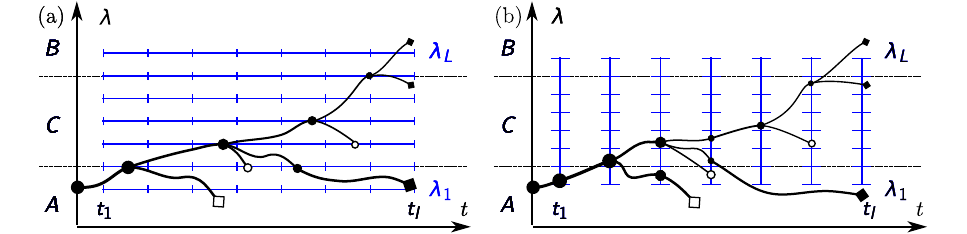}
\par\end{centering}

\caption{\label{fig:alternative_setups}In the $\lambda$-if setup (a),
  branching/pruning events are triggered by the crossing of levels
  $\{\lambda_{l}\}$ of the progress coordinate. Since in this case, the
  interface crossing flux $j_{li}$ corresponds to the flux of probability along
  $\lambda$, the $\lambda$-if setup is naturally
  suited for measuring fluxes and time-dependent transition propensities
  $q_{A\lambda}(t)$. In the $t$-if setup (b), branching/pruning is triggered at
  a set of times $\{t_{i}\}$. Here, the interface crossing flux $j_{li}$
  provides a measure of the local probability density, 
  making the $t$-if setup naturally suited for sampling $p(\lambda,t)$.}
\end{figure*}

In order to extend the FFS algorithm to non-stationary systems, we trigger
branching\slash{}pruning moves whenever a trajectory crosses an interface.
Since NS-FFS deals with systems whose dynamics are intrinsically
time-dependent, the region of interest
$\roi=[\lambda_{1},\lambda_{L}]\times[t_{1},t_{I}]$ is two dimensional. To
achieve uniform sampling of trajectories in both $\lambda$ and time, the
branching\slash{}pruning rule needs to depend on both these coordinates. To
accommodate this, we define a set of interfaces as level sets in one of the
coordinates ($\lambda$ or $t$), and partition each interface into a set of
bins, which are intervals along the other coordinate.
The region of interest $\roi$ is thus covered by a two-dimensional grid of
subdivisions (see Fig.~\ref{fig:alternative_setups}). The interfaces are used
to trigger the branching\slash{}pruning moves; the bins are used to determine
the target child number $\bar{n}$ for these moves.

The most direct generalization of FFS arises when interfaces are placed at a
set of $\lambda$-levels, and subdivided into time-bins
(Fig.~\ref{fig:alternative_setups}a).  The interface-bin grid is then given by
the sets $B_{li}=\{(x,t)|\lambda(x,t)=\lambda_{l},t_{i}\leq t<t_{i+1}\}$ where
bin
 $B_{li}$ refers to the $i$-th time bin on the $l$-th $\lambda$-interface.
This interface arrangement will be referred to as `$\lambda$-if'
($\lambda$-interfaces).
% In this setup, one can decide to either count interface crossings in both
% directions, or only the crossings in the forward direction from $A$ or $B$
% (or even only those crossings from $B$ to $A$). In either case,
One can then measure the total probability weight that has passed through bin
$B_{li}$, $H_{li}=\sum_{a=1}^{N_{li}}w_{a}$ where $N_{li}$ is the running
number of trajectories with weights $\{w_{a}\}$ that have reached $B_{li}$.
The \emph{crossing flux} per tree is
\begin{equation}
j_{li}=H_{li}/S,\label{eq:fluxest}
\end{equation}
where $S$ is the running number of sampled trees.
In computing $H_{li}$, one can choose either to count crossings only in the
forward direction (increasing $\lambda$), or in both directions, depending on
the system property of interest. When counting only forward
crossings, the quantity $\E{j_{li}}$ measures the forward probability flux
across $B_{li}$; formally
\begin{eqnarray}
\E{j_{li}}&=&\int_{t_{i}}^{t_{i+1}}
\E{\delta(\lambda(t)-\lambda_{l})\dot{\lambda}\theta(\dot{\lambda})}\dd t,
\label{eq:jli-lif}
\end{eqnarray}
where $\theta$ is the Heaviside step function, and
$\dot{\lambda}\theta(\dot{\lambda})$ is the forward probability flux. 
Although $j_{li}$ is strictly speaking a (unitless) crossing
probability, it is proportional to the forward probability flux averaged over
the bin $B_{li}$, which justifies the name `crossing flux'. Importantly, when
an absorbing boundary condition is imposed at the last interface $\lambda_L$
which is located beyond the top of the barrier, then
\begin{equation}\label{eq:jLiabs}
\E{j_{Li}} = \int_{t_{i}}^{t_{i+1}}q_{AB}(t) \dd t
\end{equation} 
where $q_{AB}$ is the first-passage time probability density, or exit
flux, into $B$. This makes the $\lambda$-if setup naturally suited for
estimating exit fluxes.

Alternatively, one may interchange the roles of time and  the progress
coordinate, by placing interfaces at a set of time points $t_{i}$, and
subdividing them into bins along the $\lambda-$direction
(Fig.~\ref{fig:alternative_setups}b). We call this interface arrangement
`$t$-if' (time-interfaces). The interface-bin grid is now described by
$B_{li}=\{(x,t)|\lambda_{l}\leq\lambda(x,t)<\lambda_{l+1},t=t_{i}\}$;
bin $B_{li}$ refers to the $l$-th $\lambda$ bin on the $i$-th time interface.
Crossing fluxes are still defined according to Eq.~\ref{eq:fluxest}, but
these have a different meaning in the $t$-if setup: $j_{li}$ simply estimates
the probability to find the system in the interval
$(\lambda_{l},\lambda_{l+1})$ at time $t_{i}$:
\begin{equation}
\E{j_{li}}=\int_{\lambda_{l}}^{\lambda_{l+1}}\E{\delta(\lambda(t_{i})
-\lambda)}\dd\lambda=\int_{\lambda_{l}}^{\lambda_{l+1}}p(\lambda,t_{i})
\dd\lambda.\label{eq:jli-tif}
\end{equation}
Thus the  $t$-if setup lends itself naturally to estimating
densities or potentials of mean force.

In both the $\lambda$-if and $t$-if setups, the branching\slash{}pruning rules
are set up to ensure uniform sampling of $j_{li}$. The $\lambda$-if setup
leads to uniform sampling of (forward) fluxes, while the $t$-if setup
implements uniform sampling of phase space densities. We stress however that
either setup could be used to measure either quantity. The relative
efficiencies of the two methods depend on the quantity used for biasing but
also on more technical aspects, as discussed below.

\subsection{Sampling with uniform error}\label{sub:equalize}

The final ingredient in the algorithm is the rule for setting the child
number probability $b(n)$ for branching\slash{}pruning moves. In NS-FFS, a
target child number is set for each bin $B_{li}$, depending on the
statistics of previous crossings of that bin; the goal of the branching rule is
to sample the crossing flux $j_{li}$ through each of the 
bins $B_{li}$ with uniform relative error.

The relative error in $j_{li}$ in an NS-FFS run may be approximated as 
(see app.~\ref{sub:Variance-of-weighted})
\begin{equation}
\frac{\E{\delta j_{li}^{2}}}{\E{j_{li}}^{2}}\simeq\frac{\alpha_{N}}{\E{N_{li}}}
\left[1+\alpha_{w}\frac{\E{\delta w_{a}^{2}}}{\E{w_{a}}^{2}}\right].
\label{eq:shortsamplingvar}
\end{equation}
Here $w_a$ denotes the (stochastic) weight of a trajectory reaching $B_{li}$,
and averages and variances refer to an ensemble of NS-FFS runs with $S$ trees.
The constants $\alpha_{N},\alpha_{w}$ approach unity in the `ideal' case where
trajectories which reach $B_{li}$ are uncorrelated.  
This expression shows that the error is
controlled by the number $N_{li}$ of trajectories that cross the bin, with an 
extra contribution arising from the spread in their weights $\E{\delta
w_{a}^{2}}/\E{w_a}^2$. Thus, to obtain a uniform relative error in $j_{li}$
(for a given total computational cost $\propto\sum N_{li}$), the branching
rule needs to equalize the number of trajectories reaching each bin, while
keeping the distribution of trajectory weights within each bin sharply peaked.
In NS-FFS, these requirements are met by using a  somewhat simplified version
of the flatPERM rule \cite{prellberg04}:

\begin{algorithm}[H]
\begin{enumerate}
\item Calculate the target child number as $\bar{n}=w_{a}/j_{li}$ where
$w_{a}$ is the weight of the incoming trajectory, and $j_{li}$ is
the current flux estimate.\label{enu:Calculate-the-target}
\item Set the child number probabilities
\[
b(n)=\delta_{n\lceil\bar{n}\rceil}(\bar{n}-\lfloor\bar{n}\rfloor)+
\delta_{n\lfloor\bar{n}\rfloor}(\lceil\bar{n}\rceil-\bar{n}),
\]
where $\lceil\cdot\rceil$ and $\lfloor\cdot\rfloor$ denote the ceiling
and floor functions, respectively.\label{enu:Set-the-child}
\item Draw a child number 
$n\in\{\lfloor\bar n\rfloor, \lceil\bar n\rceil\}$ from $b$. 
If $n>0$, set all child branch weights to $w'\leftarrow j_{li}$. 
\label{enu:Set-the-nonzero}
\end{enumerate}
\caption{\label{alg:Branching-rule}Branching rule for bin $B_{li}$}
\end{algorithm}
It is easily verified that the number of children is on average $\E
n=\bar{n}=w_a/j_{li}$. If $\bar{n}<1$, pruning occurs with probability
$1-\bar{n}$. The weight $w'$ of the children (if any) is given by the parent
weight $w_a$ multiplied by the reweighting factor $r=w'/w_{a}=1/\bar{n}$,
which is independent of $n$. It is easy to see that the condition for unbiased
statistics, Eq.~\ref{eq:unbiased}, is satisfied by this
branching\slash{}pruning rule.

The branching rule, algorithm~\ref{alg:Branching-rule}, produces a uniform
error in the crossing flux estimate $j_{li}$, since it tends to equalize
counts between bins and minimize the weight variance within a bin. To see this,
consider an NS-FFS simulation which is in steady state, i.e.~after 
the crossing flux estimates $j_{li}$ have converged to their
average values $\E{j_{li}}$.
In this situation the branching rule assigns each
trajectory leaving bin $B_{li}$ the `perfect' weight $w\to
w_{li}^{\infty}=\E{j_{li}}$. Since all child trajectories are assigned this
same weight, indeed the variance of trajectory weights leaving $B_{li}$ is minimized 
(in the ideal case it is zero). We also note that the weight
$w_{li}^{\infty}$ is equal to the system's intrinsic probability to cross bin
$B_{li}$.  It follows that 
on average exactly one trajectory per tree (with weight
$w_{li}^{\infty}$) will emanate from bin $B_{li}$  -- or equivalently
$N_{li}/S\times j_{li}\to 1\times\E{j_{li}}$ as the simulation converges. Thus
NS-FFS achieves uniform sampling of the region $\roi$, with on average an equal
number of trajectories crossing each bin, on each interface.

\subsection{The NS-FFS algorithm}\label{sub:The-NS-FFS-algorithm}

Combining the above ingredients, we now describe the full NS-FFS algorithm. To
set up the simulation, one needs to generate initial
configurations at time $t=0$, according to an initial
 distribution $\rho(x_0)$ (which need not be known explicitly as a function of
 $x_0$);
these could simply be generated via a brute-force simulation in the $A$ state.
Next, one identifies a suitable progress coordinate $\lambda$, and places a set
of interfaces, subdivided into bins $\{B_{li}\}$, over the region $\roi$ of
phase-time space of interest, according to either the $\lambda$-if or the
$t$-if setup.
One also specifies a dynamic range for reweighting, by setting minimum and
maximum trajectory weights $(w\ts{min},w\ts{max})$ (these enhance convergence,
see below).
The simulation then proceeds according to algorithm~\ref{alg:The-algorithm}:

\begin{algorithm}[H]
\begin{itemize}
\item [\it{Init.}] Define a histogram of total crossing weights $H_{li}$ and set
$H_{li}\leftarrow0$ for all $l,i$. Set a tree counter $S\leftarrow0$. Define a
queue of pending trajectories $G$ and set $G\leftarrow\{\}$. 
\label{enu:Place-a-set} 
\item [\it{Run}] Iterate the following steps, until the desired accuracy is
reached:\label{enu:Until}

\begin{enumerate}
\item Start a new trajectory at $t=0$, from a new initial state $x_{0}$.
Insert the trajectory into $G$, and assign an initial weight $w\leftarrow1$. 
Increment the tree counter, $S\leftarrow S+1$.
\item While $G$ is non-empty, iterate:\label{enu:non-empty} 

\begin{enumerate}
\item Pick and remove a trajectory from $G$\label{enu:pick-one-of}.
\item Propagate the trajectory forward in time while recording any observables
of interest, weighted with $w$, until either the final time $T$ is reached,
or an interface is crossed. In the latter
case:\label{enu:Propagate-the-trajectory}

\begin{enumerate}
\item Determine the bin $B_{li}$ which was crossed and increment $H_{li}$ by
the current trajectory weight $w$.\label{enu:increment-histogram}
\item If $w\in(w\ts{min},w\ts{max}),$ carry out a
branching\slash{}pruning move: draw a child number $n$ and set the child weight
$w'$ according to algorithm~\ref{alg:Branching-rule}.
Otherwise, set $n\leftarrow1$ and $w'\leftarrow w$. \label{enu:Draw-a-child}
\item Generate $n$ child trajectories and insert them into the queue $G$ for 
further propagation.\label{enu:Create--child}
\end{enumerate}
\end{enumerate}
\end{enumerate}
\end{itemize}
\caption{\label{alg:The-algorithm}NS-FFS}

\end{algorithm}

When setting up the region of interest $\roi$, clearly the transition region
should be covered to enhance transition paths. It is equally important to let
$\roi$ extend well into the metastable basins; this allows for pruning of
trajectories which would otherwise accumulate in the basins, degrading
performance. Note also that by default, trajectories are not terminated when
they leave $\roi$ before they reach the final time $T$, ensuring that they may
re-enter $\roi$ and contribute at a later time. It is of course possible to
explicitly add absorbing boundaries, which may increase performance, in cases
where later reentry is not required.

The weight limits $(w\ts{min},w\ts{max})$ which appear in
Alg.~\ref{alg:The-algorithm} are not strictly necessary for the existence of a
steady state with uniform sampling of the crossing flux $j_{li}$. They do,
however, greatly enhance convergence towards it. In the initial phase of an
NS-FFS simulation, the weight histogram $H_{li}$ is sparsely populated such
that the flux estimates $j_{li}$ are subject to large fluctuations. This can
result in avalanches of correlated low-weight paths which degrade
performance. For the method to be useful, it is necessary to have an effective
way of controlling these bursts. Among a number of possible remedies including
negative feedback control of tree size, branching thresholds, or explicit
flat-histogram branching based on number densities \cite{prellberg04}, we found
weight limits to be particularly simple and very effective. In practice, the
reasonable rule-of-thumb to choose $w\ts{min}\lesssim\min_{l,i}\{\E{j_{li}}\}$
and $w\ts{max}\gtrsim1$  was found to work well.

The output of algorithm~\ref{alg:The-algorithm} consists of weighted trees of
trajectories in which all trajectory segments end either at an interface
(branching/pruning point) or at the final time $T$ (completion) (Note that
a simulation may be stopped only after a full tree is finished). Trees may be
generated depth-first (children before sisters), or breadth-first (sisters
before children), depending on whether the queue $G$ in
Alg.~\ref{alg:The-algorithm} is of the last-in-first-out or first-in-first-out
type. We found no significant difference in performance between depth-first and
breadth-first traversal, in contrast to other reports for the case of PERM
\cite{hsu11}. Trajectory trees may also be written to disk in a recursive data
structure for offline analysis.

\section{Applications }\label{sec:Applications}

\subsection{Crossing of a linear barrier}

\begin{figure}
\begin{centering}
\includegraphics[width=8cm]%
{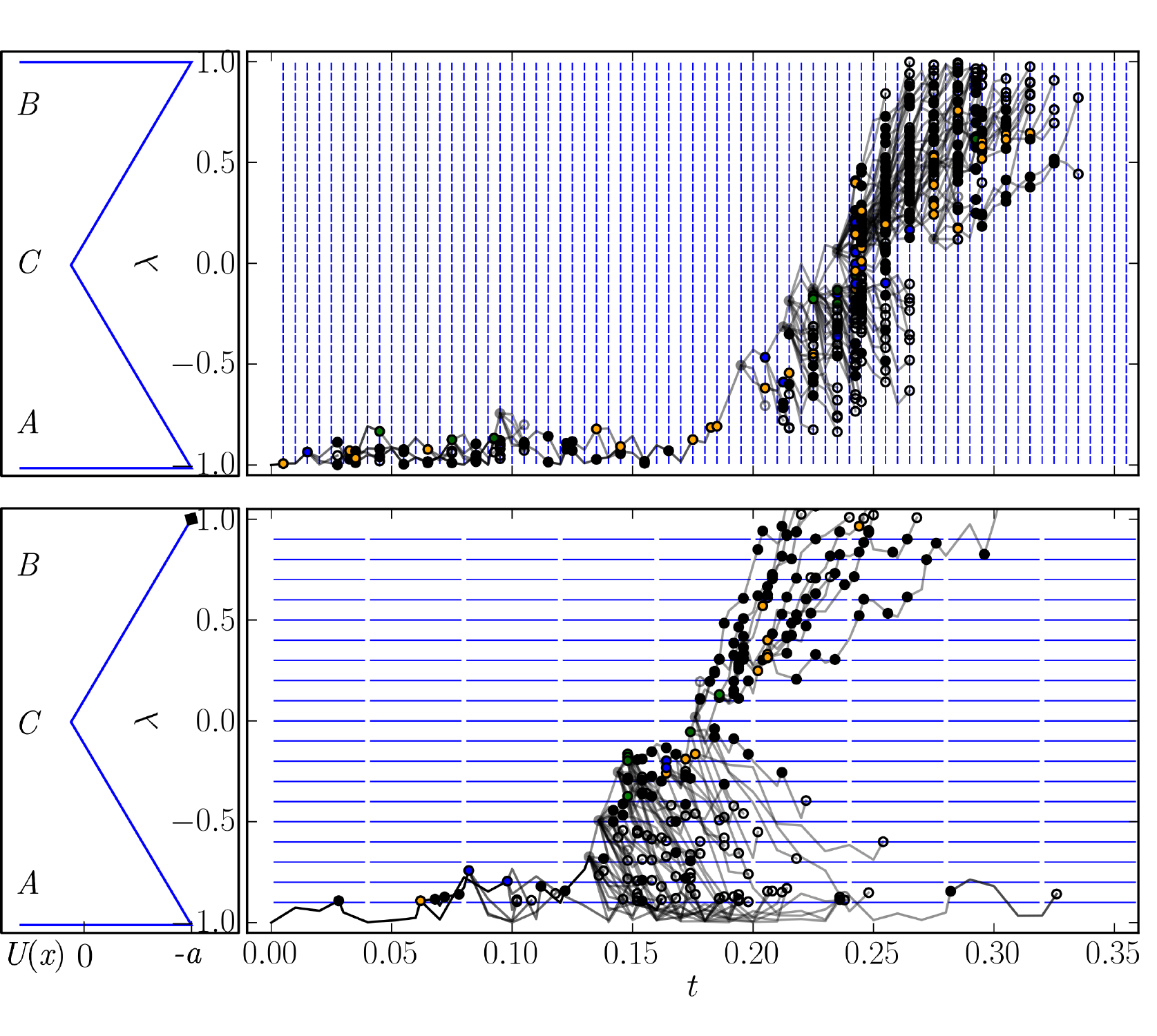} \par\end{centering}

\caption{\label{fig:barrier_pot_tree}Linear ramp potential $U$ (left) and
the first 500 branches of path trees generated during NS-FFS simulations
using the $t$-if setup (top, reflecting boundaries) and the $\lambda-$if
setup (bottom, reflecting\slash{}absorbing boundaries). Although the total
simulated time that is shown, is $\tau_{sim}\lesssim6\ll\tau\ts{rxn}$,
crossing events are successfully generated. Bins $B_{li}$ are depicted
in blue, branch weight is indicated by line shading and branch points
are shown as circles. }
\end{figure}

As a simple model for rare barrier crossing events, we consider overdamped
Brownian motion in a linear double-ramp potential $U=-a|x|$ with boundaries at
$x=\pm1$ and barrier height $a>0$, see Fig.~\ref{fig:barrier_pot_tree}. We
consider two systems, which differ in their boundary conditions: one with
reflecting boundaries at both $x=-1$ and $x=1$, meaning that probability is
conserved; another one with a reflecting boundary condition at $x=-1$ and an
absorbing one at $x=1$.  The two systems will be referred to as the
reflecting\slash{}reflecting or reflecting\slash{}absorbing systems,
respectively. This choice of model potential allows for an exact calculation of
the Green's function, so that the correctness of our simulations can be
assessed directly.

Particles are injected at $x=-1$, $t=0$ and diffuse according to
the overdamped Langevin equation
\begin{equation}
\dot{x}=-D\partial_{x}U+\xi.\label{eq:langevin}
\end{equation}
We use thermal units where $k_{B}T=1$, so that $\xi$ represents Gaussian
white noise with $\E{\xi(t)}=0,$ $\E{\xi(t)\xi(t')}=2D\delta(t-t')$.
In these units, the diffusion constant $D$ is equal to the mobility
coefficient, and $a$ has units of inverse length. 

In this  barrier escape problem both the time scale $\tau_{A}$ for
equilibration in region $A$ for $x\gtrsim-1$ and the crossing time $\tau_{C}$,
are much faster than the waiting time between crossings
$\tau_{AB}=\tau_{BA}=2\tau\ts{rxn}$ where $\tau\ts{rxn}$ is the global
relaxation time of the system. The Laplace-transformed probability density
$p(x,s)=\int_{0_{-}}^{\infty}e^{-st}p(x,t)\dd t$ can be calculated exactly
using standard methods, see app.~\ref{sec:Piecewise-linear-potential}, leading
to explicit expressions for the time scales
\begin{equation}
\tau_{A}=\frac{4}{a^{2}D}\text{, }\tau_{C}=\frac{2}{aD}\text{, and
}\tau\ts{rxn}=\frac{2e^{a}}{a^{2}D}.\label{eq:barriertimescales}
\end{equation}
In the present examples, the barrier height and diffusion constant
were set to $a=15$ and $D=1$, respectively, such that
$(\tau_{A},\tau_{C},\tau\ts{rxn})=(0.02,0.13,2.9\times10^{4})$,
respectively. Moreover, the full probability density $p(x,t)$, and
for the reflecting\slash{}absorbing case, the exit propensity $j(x,t)$,
could be obtained as functions of time using an efficient numerical contour
integration method \cite{weideman07}.

Fig.~\ref{fig:barrier_pot_tree} depicts the potential, the interface bins, and
partial trajectory trees taken from NS-FFS simulations of this system from
$t=0$ to $t=T=1$. The region of interest was taken to be
$\roi=[-1,1]\times[0,1]$ in $(x,t)$; the progress coordinate $\lambda$ was
defined trivially as $\lambda\equiv x$.

\begin{figure}
\begin{centering}
\includegraphics[width=8cm]%
{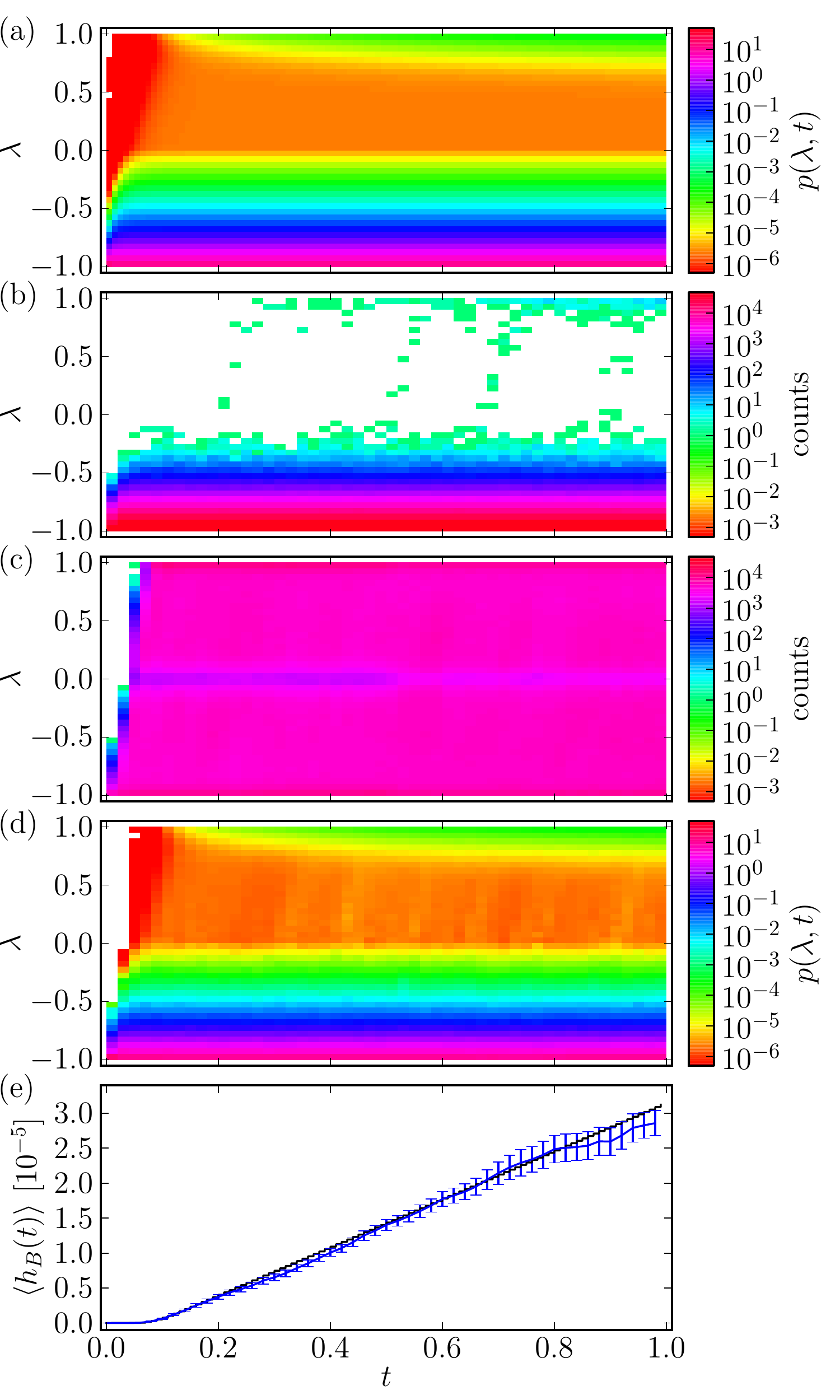} \par\end{centering}

\caption{\label{fig:barrierdensities}Time-dependent density  $p(\lambda,t)$ for
the linear ramp potential with initial condition
$p(\lambda,0)=\delta(\lambda+1)$, with reflecting boundary conditions. (a)
numerically exact solution; (b) brute force sampling.
In the $t$-if NS-FFS run, raw counts (c) cover $\roi$ almost uniformly while
weighted counts (d) reproduce the exact density. Taking the $B$ state boundary
to be $q_{B}=0.5$, the occupation of the $B$ state (d) fits a linear growth
model with slope $k_{AB}=(3.41\pm.03)\times10^{-5}$ and delay $0.099\pm.003$
(blue, as extracted from NS-FFS; black, exact solution).
Total simulated time was $10^{5}\simeq3.4\tau\ts{rxn}$. Counts refer to the
histogram bins of size $(\Delta x,\Delta t)=(.05,.01)$ used in this figure.}
\end{figure}

We first study the probability density $p(\lambda,t)$ in the 
reflecting\slash{}reflecting system,
which relaxes towards the Boltzmann distribution $\propto e^{-U}$
over times longer than $\tau\ts{rxn}$. The system was simulated using
the $t$-if setup which is most natural for measuring $p(\lambda,t)$, since
the branching factors are controlled by the local density. $I=199$
interfaces were placed across $\roi$, at regular time intervals 
(from $t=0$ to 1), and partitioned into $L=40$ equal-sized bins each 
(from $\lambda=-1$ to $1$). 
In Fig.~\ref{fig:barrierdensities} we compare $p(\lambda,t)$ as obtained as via
exact analytical calculation, brute-force and NS-FFS simulation.
While the probability density $p$ varies over 8 decades  within the region
$\roi$, the sampling density in NS-FFS is constant within a factor
of 5 (Fig.~\ref{fig:barrierdensities}c). 
After reweighting, $p$ is correctly reproduced throughout $\roi$ (even where
it is very small; Fig.~\ref{fig:barrierdensities}d). 
This was achieved within a simulation time that would
generate only a handful of transition paths in a brute force simulation 
(Fig.~\ref{fig:barrierdensities}b).
The region around the cusp of the potential is somewhat under-sampled.
This is due to the fact that the force in our model is discontinuous at $x=0$; 
to achieve complete sampling uniformity in this region, one 
would require a bin width on the order of the length
scale of variation of the potential. 

\begin{figure}
\begin{centering}
\includegraphics[width=8cm]%
{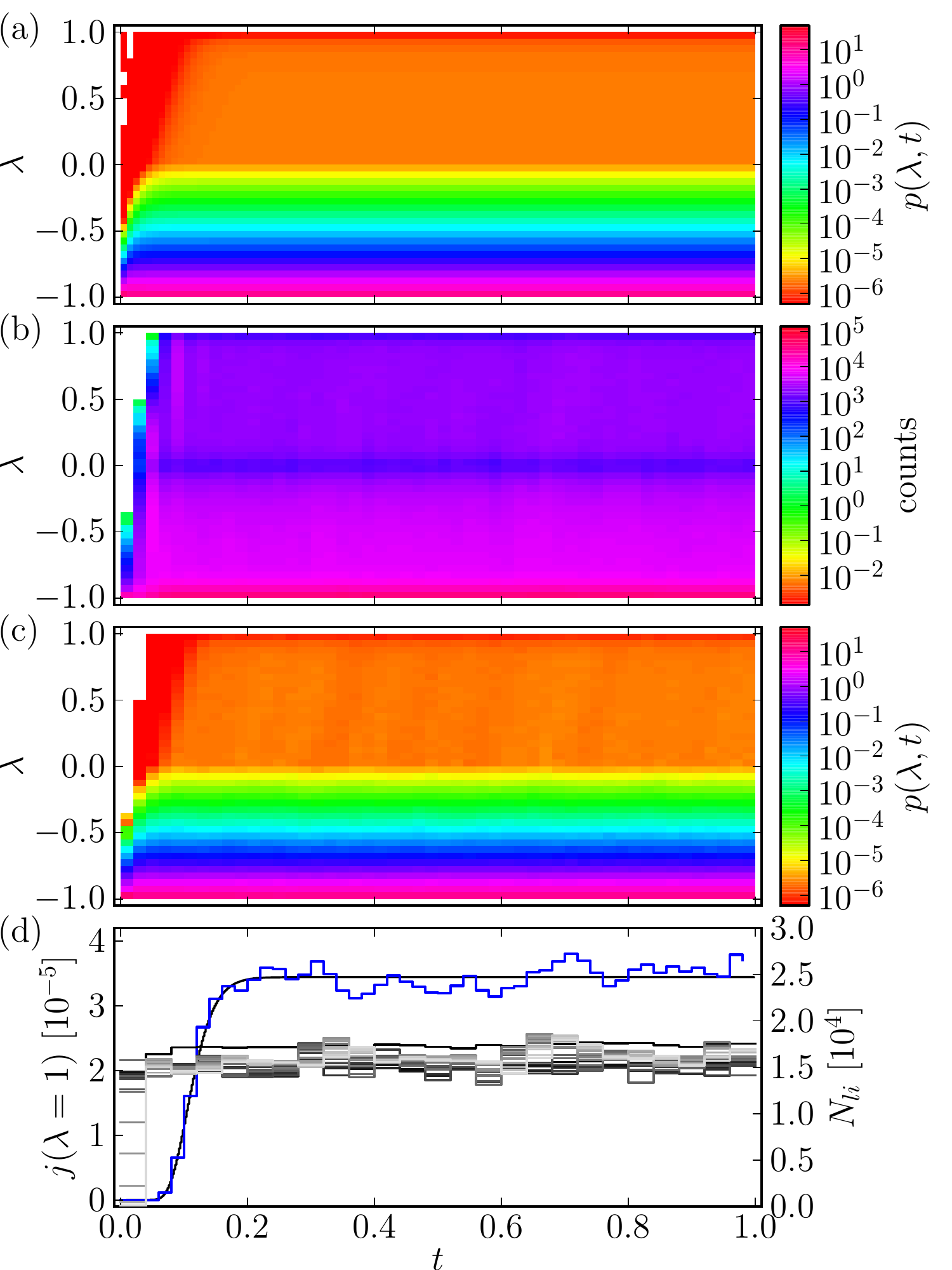} \par\end{centering}

\caption{\label{fig:barrierabsdensities}Time-dependent probability density (as
in Fig.~\ref{fig:barrierdensities}) and exit flux, but for
reflecting\slash{}absorbing boundary conditions. The exact density $p$ is shown
in (a). Counts generated in a $\lambda$-if NS-FFS run show roughly uniform
sampling (b, see text). Weighted counts (c) reproduce the exact density over
the full dynamic range. Unweighted number fluxes of trajectories emanating from
each bin (d, right axis) for all 19 $\lambda$-interfaces (gray value increasing
with $\lambda)$ collapse, indicating uniform number flux. As a consequence the
exit probability flux $j$ over $x=1$ (d, left axis) is sampled uniformly
including its low-probability onset (black, exact solution; blue, NS-FFS). An
estimate of the stationary exit flux from these data for $t>.24$ gives
$j=3.424(\pm.02)\times10^{-5}$; the exact value is
$j_{\infty}=\tau_{AB}^{-1}=3.44\times10^{-5}$. }
\end{figure}

We next consider the time-dependent exit flux through the absorbing boundary
in the reflecting\slash{}absorbing system. For this NS-FFS calculation, we used 
the $\lambda$-if setup; this setup is natural since here the
sampling bias is based on crossing fluxes over the $\lambda$-interfaces,
and the last crossing flux coincides with the observable of interest.
$L=19$ $\lambda$-interfaces were placed at regular intervals and
partitioned into $I=50$ time-bins each; only crossings in the positive 
positive $\lambda$-direction counted towards  $H_{li}$.
Fig.~\ref{fig:barrierabsdensities} shows the probability density $p(\lambda,t)$
computed using the exact solution (a) and NS-FFS (c) as well as the unweighted
crossing fluxes and the reweighted exit flux (d) for this system.

As expected, the number fluxes of trajectories emanating from each bin in the
positive lambda direction (d, right axis) superimpose in a narrow
band, confirming that NS-FFS indeed produces uniform sampling across $\roi$.
Only the earliest bins at the farthest interfaces are visited less often, since
the system dynamics does not allow them to be reached in the required time with 
sufficient probability.

In contrast, the number density of trajectories over $\lambda, t$ (b) is
roughly uniform over $\roi$, but does exhibit a systematic bias favoring the
$A$ state. This can be understood as follows. The branching rule in the chosen
$\lambda$-if setup biases towards uniform number flux in the forward
$\lambda$-direction, not uniform density.
While in the region $x>0$ trajectories spontaneously move in positive flux
direction, in the region $x<0$, trajectories tend to drift downhill, against
the positive flux direction. In this region, the $\lambda$-if NS-FFS simulation
maintains a uniform population of uphill trajectories by proliferating. The
total (uphill+downhill) trajectory density is thus increased in the region
$x<0$.

This observation clearly demonstrates the difference in sampling biases between
the $t$-if and $\lambda$-if setups: the former generates a uniform density
while the latter generates uniform fluxes in the $\lambda$-direction. 
Nevertheless, as Fig.~\ref{fig:barrierabsdensities}b shows, in practice the
methods provide almost uniform sampling of both forward flux and total number
density, so that it is certainly possible to sample fluxes using the $t$-if
setup and densities with the $\lambda$-if setup, without dramatic
loss in performance.

In this context it is worth noting that depending on the observable to be
estimated, there is the additional freedom of setting up the simulation to
measure either density or exit flux. For instance, the slope in
Fig.~\ref{fig:barrierdensities}(e) and the plateau value of the exit flux in
Fig.~\ref{fig:barrierabsdensities}(d) coincide, even though trajectories may
re-cross the barrier from $B$ to $A$ in the reflecting\slash{}reflecting
system, but not in the reflecting\slash{}absorbing system.  This is in
agreement with the exact expressions in the limit $t\ll\tau\ts{rxn}$; indeed,
as long as $t\ll\tau\ts{rxn}$, the occupation of the $B$ state is negligibly
small so that back-crossings occur with a probability even much smaller than
forward crossings. Effectively, the $B$ state initially appears as absorbing
even in the purely reflecting system. Thus, either simulation may be used to
measure the rate constant for this system.

\begin{figure}
\begin{centering}
\includegraphics[width=8cm]%\
{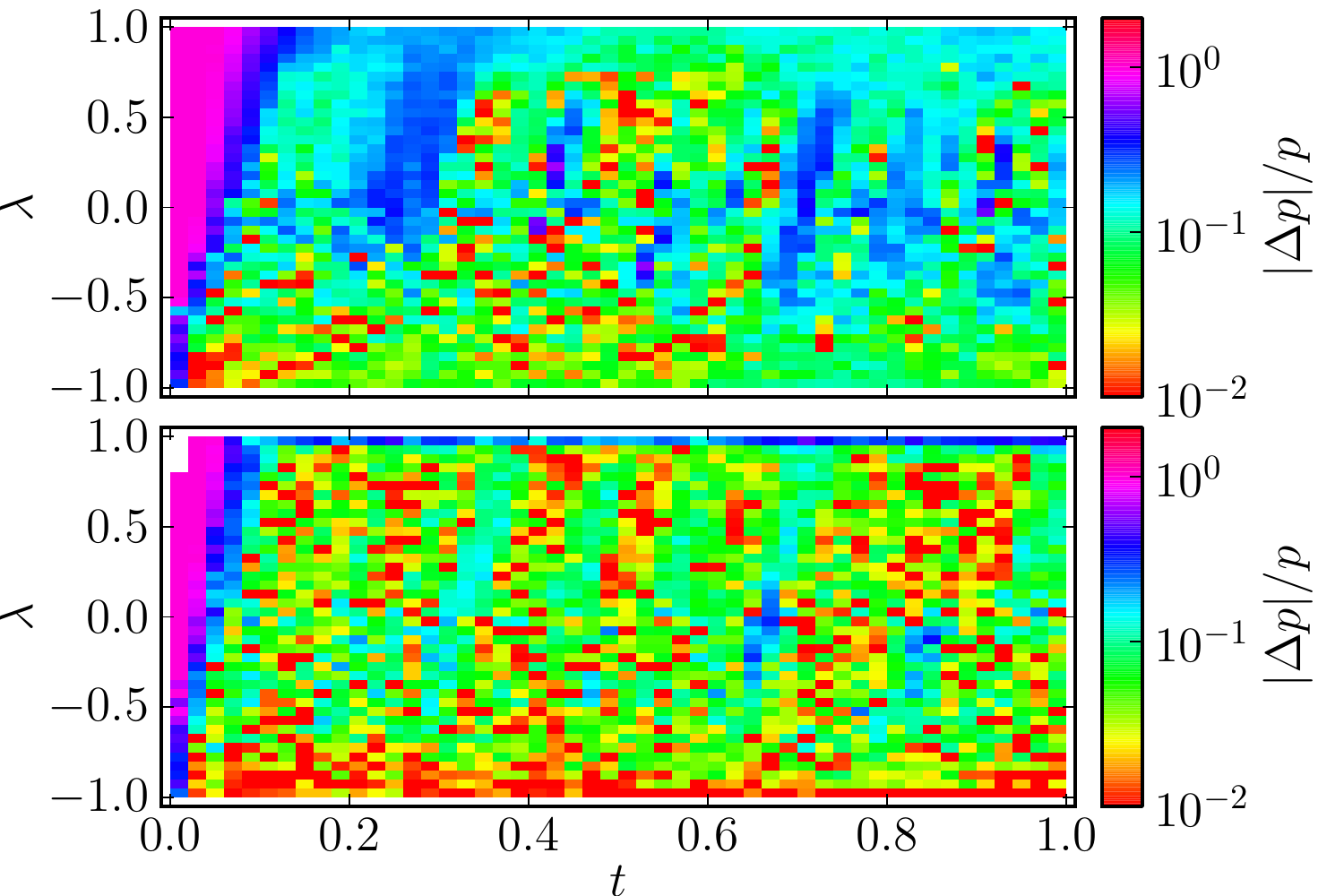} \par\end{centering}

\caption{\label{fig:barriererrors}Relative errors corresponding to the 
probability densities shown in figs.~\ref{fig:barrierdensities} and 
\ref{fig:barrierabsdensities} for the $t$-if
and $\lambda$-if (top and bottom, respectively). Relative errors are
computed as absolute differences between sampled and exact densities,
normalized by the exact density.}
\end{figure}

The relative errors in the estimated probability density $|\Delta
p|/p=|p\ts{sim}(\lambda,t)-p\ts{exact}(\lambda,t)|/p\ts{exact}(\lambda,t)$
shown in Fig.~\ref{fig:barriererrors}, further illustrate the uniform sampling
over $\roi$ generated by NS-FFS. The relative error scatters uniformly over
$\roi$ and in particular, does not scale with $p^{-1/2}$ as would be the case
for brute-force sampling. Larger errors remain only in the fringes of the
accessible region,  where fewer trajectories are sampled.
The residual stripe pattern in the $t$-if case carries the signature of
correlated trajectories originating at the cusp of the potential; the
undersampling right at the cusp which causes this can be considered a
pathological feature of the force discontinuity in our model.
Notice that although the $\lambda$-if setup equalizes positive fluxes rather
than number density, the probability density of trajectories nevertheless
exhibits uniform error in this example, despite the weak asymmetry across
$\roi$ in the number of trajectories sampled, visible in
Fig.~\ref{fig:barrierabsdensities}b.

\subsection{Genetic toggle switch}\label{sub:Toggle-switch}

As an intrinsically non-equilibrium example, we next consider a bistable gene
regulatory network which can be seen as a simplified version of the
$\lambda$-phage genetic switch. This `toggle switch' consists of two genes that
mutually repress each other. In the `exclusive' variant considered here, the
two genes, which produce proteins A and B respectively, share a common DNA
operator region $\mathrm{O}$, such that when the dimer ${\rm A}_{2}$ is bound
to the operator, protein B cannot be produced, and vice versa. This model is
discussed in detail in refs~\cite{warren04a,warren05}.

In this simplified model, production of proteins is represented by a Poisson
process. The model consists of the symmetric reaction set:
\begin{equation}
\begin{array}{ccc}
\mathrm{O} & \overset{k}{\to} & \mathrm{O}+\mathrm{A}\\
\mathrm{A} & \overset{\mu}{\to} & \varnothing\\
\mathrm{A}+\mathrm{A} & 
  \overset{k\ts f}{\underset{k\ts b}{\rightleftharpoons}} & \mathrm{A}_{2}\\
\mathrm{O}+\mathrm{A}_{2} & 
  \overset{k\ts{on}}{\underset{k\ts{off}}{\rightleftharpoons}} &
\mathrm{O}\mathrm{A}_{2}\\
\mathrm{O}\mathrm{A}_{2} & 
  \overset{k}{\to} & \mathrm{O}\mathrm{A}_{2}+\mathrm{A}
\end{array}\quad\begin{array}{ccc}
\mathrm{O} & \overset{k}{\to} & \mathrm{O}+\mathrm{B}\\
\mathrm{B} & \overset{\mu}{\to} & \varnothing\\
\mathrm{B}+\mathrm{B} & 
  \overset{k\ts f}{\underset{k\ts b}{\rightleftharpoons}} & \mathrm{B}_{2}\\
\mathrm{O}+\mathrm{B}_{2} & 
  \overset{k\ts{on}}{\underset{k\ts{off}}{\rightleftharpoons}} &
\mathrm{O}\mathrm{B}_{2}\\ \mathrm{O}\mathrm{B}_{2} & 
  \overset{k}{\to} & \mathrm{O}\mathrm{B}_{2}+\mathrm{B}.
\end{array}\label{eq:toggle}
\end{equation}
The fact that only protein dimers may bind to the operator site, 
and the fact that the state
$\mathrm{OA_{2}B_{2}}$ is disallowed, together make this system a robust
bistable switch \cite{warren04a}.  Each metastable state is characterized by 
an abundance of only one species, and transitions between these states occur 
on a much longer timescale than relaxation within
them. We use the same rate constants as in \cite{allen06a}: $\mu=k/4,k\ts
f=k\ts b=5k,k\ts{on}=5k\text{ and }k\ts{off}=k$, and we measure time in units
of $k^{-1}$.

A natural progress coordinate for this non-equilibrium system is given by the
difference in total monomer numbers, \[
\lambda=n_{\mathrm{B}}+2n_{\mathrm{OB}_{2}}+2n_{\mathrm{B}_{2}}-(n_{\mathrm{A}}
+2n_{\mathrm{OA}_{2}}+2n_{\mathrm{A}_{2}}).
\] We are interested in a region
$\roi=\{(\lambda,t)\in(-40,40)\times(0,10^{3})\}$ which spans both  metastable
states and the transition region. Using the $t$-if setup, we define $I=500$
equidistant time interfaces. $L=16$ $\lambda$-bins were defined with boundaries
at $\lambda=\pm$\{40, 24, 22, 18, 15, 12, 9, 4\} (These are the interface
locations used in \cite{allen06a}, augmented by bins in the basins $A,B$.) We
take the metastable states as $A=\{(x,t)|\lambda<\lambda_{A}\}$ and
$B=\{(x,t)|\lambda>\lambda_{B}\}$ where $-\lambda_{A}=\lambda_{B}=24$.

\subsubsection{Unbiased relaxation}\label{sub:Relaxation-to-stationary}

\begin{figure}
\begin{centering}
\includegraphics[width=8cm]%
{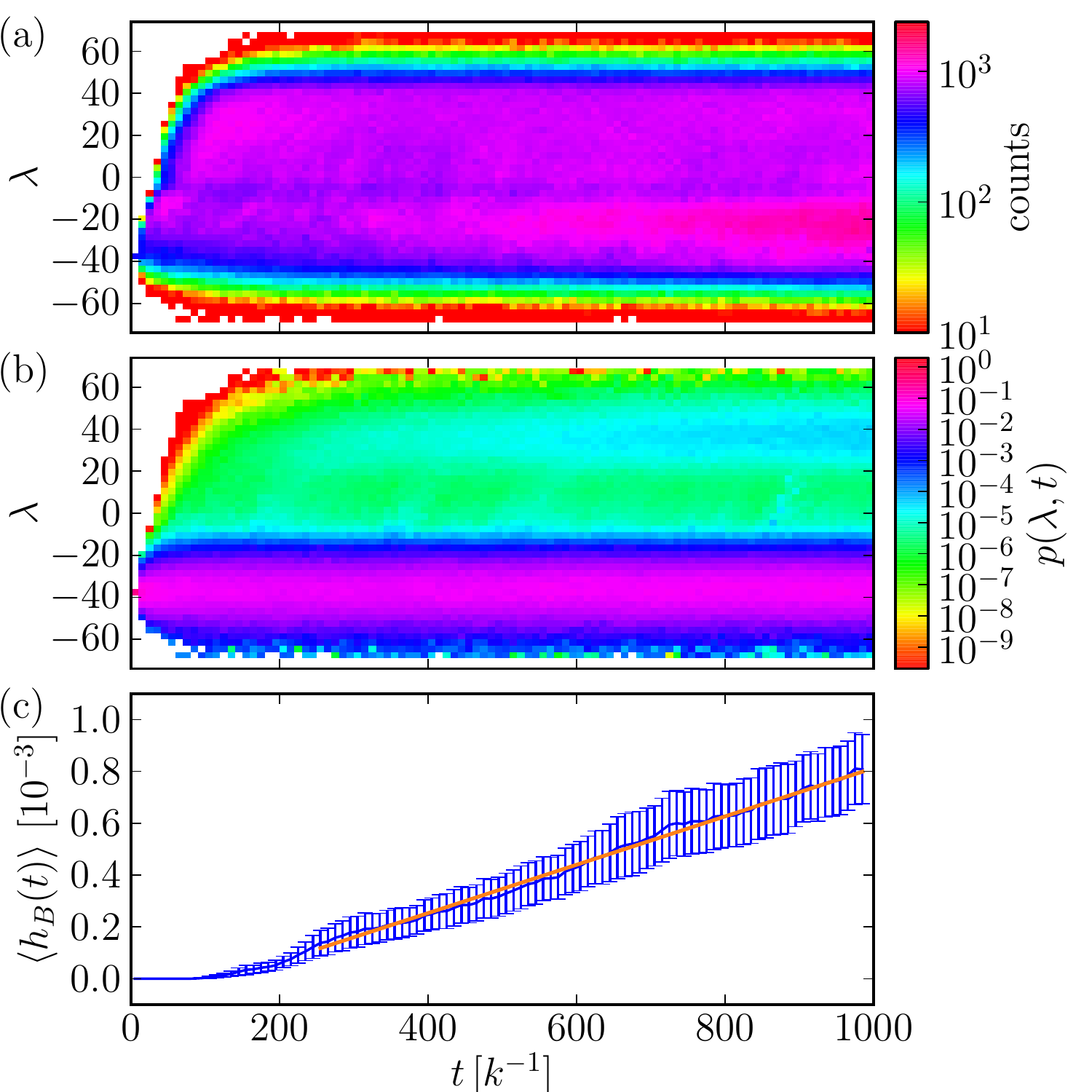} \par\end{centering}

\caption{\label{fig:togg1}Bin counts (a), probability density $p(\lambda,t)$
(b), and cumulative crossing probability $\E{h_{B}(t)}$ (c) for the toggle
switch, simulated using NS-FFS with the $t$-if setup. A linear fit for
$t>250$ is shown in orange. Error bars were generated by bootstrap resampling
from 25 independent simulation runs with total simulated time $10^{6}$ each.}
\end{figure}

Fig.~\ref{fig:togg1} shows the result of an NS-FFS simulation of this model
genetic switch from $t=0$ to $T=10^{3}$, with initial molecule numbers
fixed to $0$ except $n_{\mathrm A}=n_{\mathrm{A}_2}=10$, so that
$\lambda(0)=-30\in A$. As in the previous one-dimensional example, the region
of interest is sampled approximately uniformly in NS-FFS. Measuring the
occupancy of the $B$ state $\E{h_{B}(t)}=\E{\theta(\lambda(t)-\lambda_{B})}$,
and fitting to a delayed linear rise $(t-\tau\ts{lag})/\tau\ts{AB}$, we obtain
a lag time $\tau\ts{lag}=(129\pm5),$ and recover a waiting time for barrier
crossing $\tau\ts{AB}=(1.07\pm.01)\times10^{6}$ in accordance with the
previously measured value $1/k\ts{AB}=(1.06\pm.02)\times10^{6}$
~~\cite{allen06a}.

\subsubsection{Response to time-dependent forcing}
\label{sub:Response-to-time-dependent}

We now consider the reaction of the toggle switch to a time-dependent external
bias. This case is inspired by the phage-$\lambda$ switch in the bacterium
${\it{Escherichia\,\,coli}}$, where an increase in intracellular RecA
concentration triggers the transition from the lysogenic to the lytic phase of
the virus life cycle \cite{ptashne04}. As a simplified model for the
action of RecA we introduce a species $\mathrm{R}$ which degrades $\mathrm{A}$
monomers:
\begin{gather}
\varnothing\overset{k_{\mathrm{R}}}{\to}\mathrm{R}
  \overset{\mu_{\mathrm{R}}}{\to}\varnothing\nonumber \\
\mathrm{A}+\mathrm{R}\overset{\gamma}{\to}\mathrm{R}\label{eq:toggle control}
\end{gather}
The degradation of $\mathrm{A}$ by $\mathrm{R}$ forces the switch towards the
$B$ state; thus  $\mathrm{R}$ can be regarded as an `external force' acting
on the switch, whose strength can be measured by the steady-state bias
$\gamma n_{\mathrm{R}}^{\infty}$, where $n_{\mathrm{R}}^{\infty}=
k_{\mathrm{R}}/\mu_{\mathrm{R}}$ is the number of molecules of $\mathrm{R}$
in steady state. Relaxation of $n_{\mathrm{R}}$ towards 
$n_{\mathrm{R}}^{\infty}$ is exponential with a relaxation time
$\mu_{\mathrm{R}}^{-1}$.

\begin{figure}
\begin{centering}
\includegraphics[width=8cm]%
{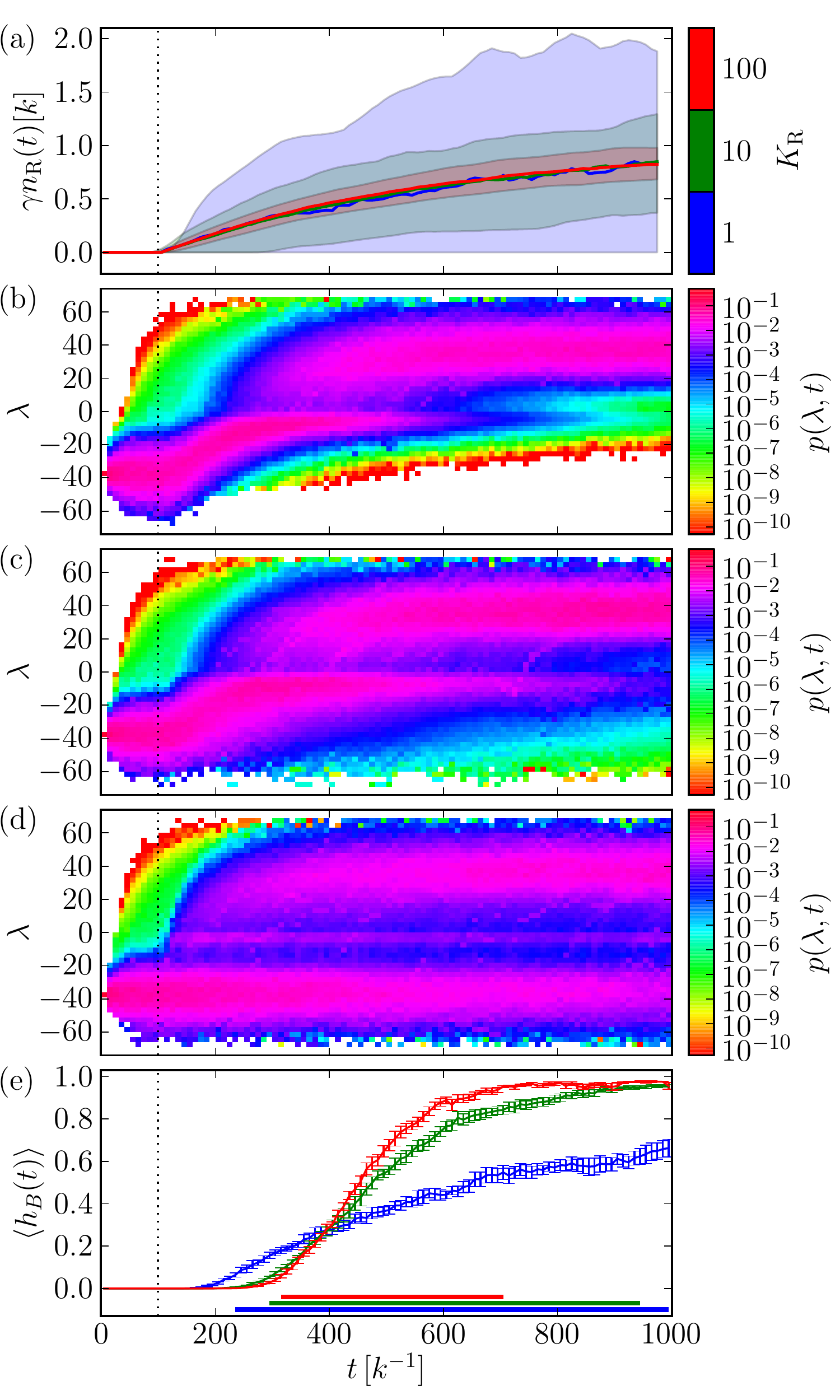}
\par
\end{centering}

\caption{\label{fig:on-switch}Response of the toggle switch to a bias of
$\mathrm{R}$ molecules, whose production begins at time $t=0$
(Eq.~\ref{eq:toggle control}).
The mean bias strength approaches $\gamma n_{\mathrm{R}}^{\infty}=k$ (a, lines)
but with different amounts of noise, corresponding to different choices for $
n_{\mathrm{R}}^{\infty}$ (a, shaded areas lie between the $5$th and $95$th
percentiles). The switch response becomes more random as the noise in the bias
increases ($n_{\mathrm{R}}^{\infty}$ decreases; b-d, respectively), and the
transition time distribution widens (e, error bars indicate the $5$th and
$95$th percentiles for the $B$ state occupation $\E{h_B(t)}$).}
\end{figure}

First, we initialize the switch in state $A$ with $n_{\mathrm{R}}=0$ copies of
$\mathrm{R}$ and switch on the production of $\mathrm{R}$. The steady state
bias $\gamma n_{\mathrm{R}}^{\infty}$ is chosen such that in steady state the
switch is fully driven to the $B$ state. The switch then flips from $A$ to $B$
with a distribution of switching times. Switching events result from favorable
fluctuations in the copy numbers of the molecules that constitute the switch
(eqs.~\ref{eq:toggle}). These are partly due to the intrinsic stochastic nature
of the switch, and partly induced by (extrinsic) fluctuations in the number of
biasing molecules $\mathrm{R}$ (eqs.~\ref{eq:toggle control}). The distribution
of switching times thus reflects both intrinsic noise of the switch and
extrinsic noise originating from fluctuations in the number of ${\mathrm R}$
molecules.
We investigated these effects by using NS-FFS to obtain the switching dynamics
as a function of the level of noise in the bias.
To modulate the latter, we varied the equilibrium copy number of $\mathrm{R}$
between $n_{\mathrm{R}}^{\infty}=1$ and $n_{\mathrm{R}}^{\infty}=100$, while
keeping the average bias $\gamma n_{\mathrm{R}}^{\infty}$ and bias time
constant $\mu_{\mathrm{R}}^{-1}\equiv500$ fixed. Therefore, on a mean-field
level, all bias protocols were kept the same. However, the individual bias
trajectories $n_{\mathrm{R}}(t)$ are markedly different: At
$n_{\mathrm{R}}^{\infty}=100$, each trajectory $n_{\mathrm{R}}(t)$ exhibits a
nearly deterministic and exponential rise in time, while at
$n_{\mathrm{R}}^{\infty}=1$, each individual trajectory $n_{\mathrm{R}}(t)$ is
a single off-on event, which is exponentially distributed in time.
Fig.~\ref{fig:on-switch}a shows the mean exponential rise of the bias protocol
and the level of fluctuations around it. Note that in a linear system
where the state occupancies are linear functions of the external forcing
history, this variation of parameters would lead to a probability of having
switched after the pulse which is independent of the pulse duration
$\mu_{\mathrm{R}}^{-1}$.

Fig.~\ref{fig:on-switch}(b-e) illustrates the switch response to a bias at
various noise levels. For a nearly deterministic bias, 
with $n_{\mathrm{R}}^{\infty}=100$,
the switch response is characterized by a gradual increase of
$\lambda$ towards a threshold, followed by a transition over the
threshold (Fig.~\ref{fig:on-switch}b). The transition times have a
relatively narrow distribution around $t=500$. Since further
increasing the expression level $n_{\mathrm{R}}^{\infty}$ does not
sharpen the switching time distribution (not shown), the result shown
in panel b corresponds to the intrinsic limit in precision of the
toggle switch at the given rate of biasing $\mu_{\mathrm{R}}$.

As the driving becomes noisier (Fig.~\ref{fig:on-switch}c,d), the switch
response exhibits a broader distribution of switching times, with both early
and late crossings, and in addition, a weak new metastable state develops at
the transition state (d). Fig.~\ref{fig:on-switch}e summarizes these results,
showing the probability that the switch has flipped, as a function of time, for
several different values of $n_{\mathrm{R}}^{\infty}$. It is clear that noise
in the driving force has an important effect on the switching trajectories.

The qualitative change in the switch flipping trajectories for
low values of $n_{\mathrm{R}}^{\infty}$ can be
understood by a picture in which the switch dynamics are enslaved to 
fluctuations in $\mathrm{R}$.  In the extreme case
$n_{\mathrm{R}}^{\infty}=1$  (Fig.~\ref{fig:on-switch}d), individual bias 
trajectories switch to full bias strength suddenly, at random
times. The switch then responds to the strong bias by rapid
flipping to the $B$ state in a stereotyped way; the time course of the 
switching event (not shown) includes a pause at the threshold
which is responsible for the zone of higher density around $\lambda=0$
(This pause originates from the fact that for $n_{\mathrm{R}}^{\infty}=1$, the
$A$ molecules are very rapidly degraded when the single $R$ molecule
becomes present, while it still takes time to produce the $\mathrm B$
molecules, which ultimately flip the switch).

We now consider the switch response to transient pulses of biasing molecules
$\mathrm{R}$. In these simulations, after the system reaches a quasi-stationary
state in region $A$, $n_{\mathrm{R}}\tsup{pulse}=10^{2}$ biasing molecules are
flushed in instantaneously. We set $k_{\mathrm{R}}=0$, so $n_{\mathrm{R}}$ then
decays to 0 over a pulse duration $\mu_{\mathrm{R}}^{-1}$ . During the pulse,
the switch is biased towards the $B$ state. To isolate the effect of different
pulse durations on the switch response, we adjust $\gamma$ such that the
integrated bias strength $\E{\int\gamma n_{\mathrm{R}}(t)\dd t}\equiv1$ remains
fixed, while changing the value of $\mu_{\mathrm{R}}$
($\mu_{\mathrm{R}}^{-1}=100,1,0.1$). This leads to
$\gamma=\mu_{\mathrm{R}}/n_{\mathrm{R}}\tsup{pulse}$.

\begin{figure}
\begin{centering}
\includegraphics[width=8cm]%
{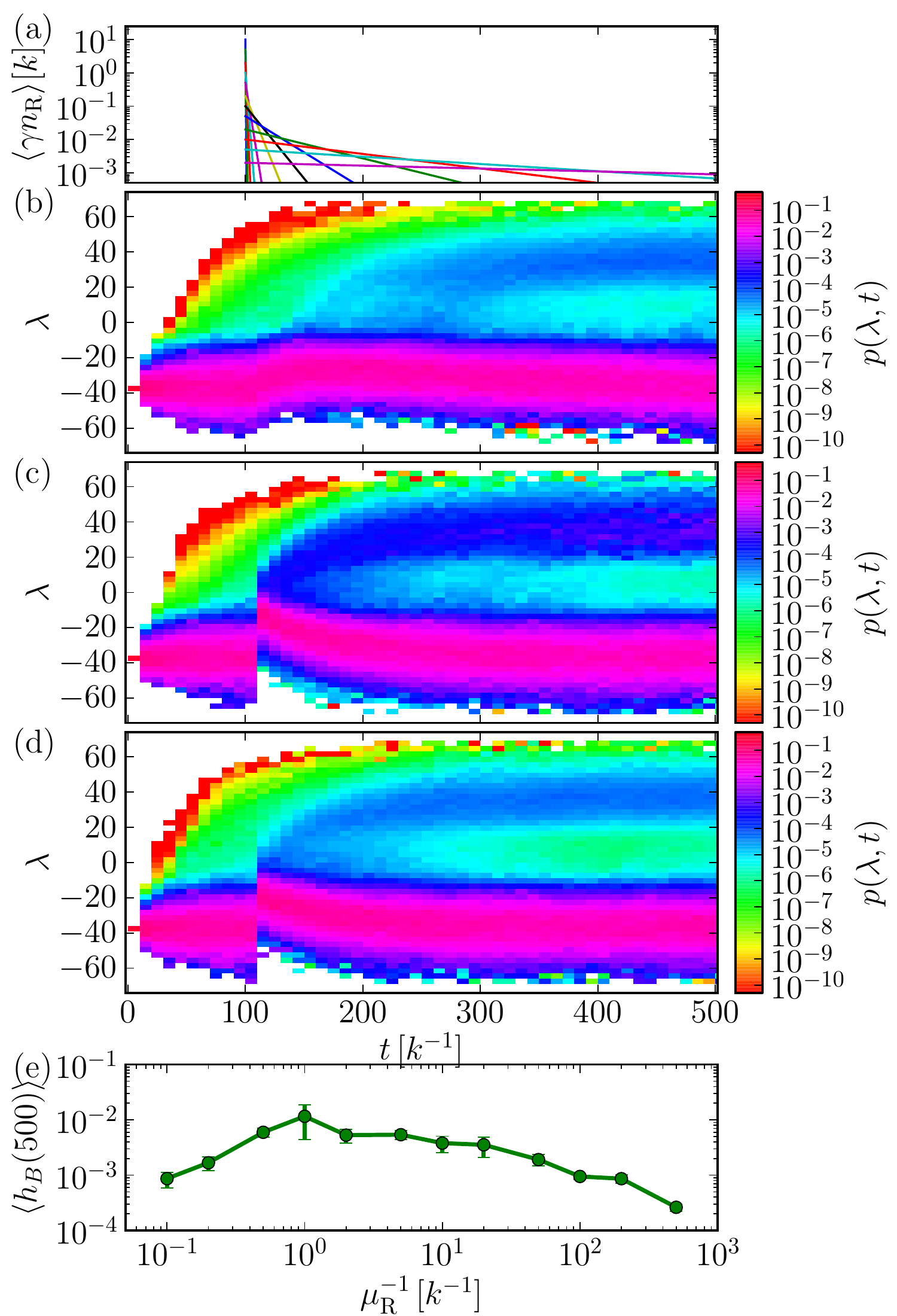} \par\end{centering}

\caption{\label{fig:pulse}NS-FFS results for the toggle switch, under the
influence of pulses of biasing molecules ${\mathrm{R}}$ of varying duration
$\mu_{\mathrm{R}}^{-1}$  but constant total efficiency
 $\E{\int\gamma n_{\mathrm{R}}(t)\dd t}\equiv1$. The time-dependent bias
 $\E{\gamma n_{\mathrm{R}}}$ is shown in (a).  The response
of the system to pulses of durations $\mu_{\mathrm{R}}^{-1}=\{100,1,0.1\}$ is
shown in (b,c,d), respectively. The crossing probability, 
taken to be the $B$ state occupation $\E{h_B(t=500)}$ exhibits a maximum at
a pulse duration around $\mu_{\mathrm{R}}^{-1}=1$ (e)}
\end{figure}

Fig.~\ref{fig:pulse} shows that the toggle switch possesses an
optimal pulse duration, even though the integrated bias strength
remains constant. For moderately long pulses, the occupation of the
$B$-state after $t=500,$ increases with decreasing pulse duration. This
can be understood in terms of a non-linear threshold behavior of the switch.
As the pulse duration $\mu_{\mathrm{R}}^{-1}$ is decreased, the initial bias
strength $\gamma n_{\mathrm{R}}\tsup{pulse}=\mu_{\mathrm{R}}$ increases,
enhancing the switching probability.
However, as the pulses become even shorter the switching probability decreases
again. This decrease is a dynamical effect:
since the switch cannot respond to changes in $n_{\mathrm{R}}$ which occur on a
timescale shorter than its own intrinsic kinetic time scale $k^{-1}$, it acts
as a low-pass filter. In this sense the toggle switch can be said to be robust
against both strong transient perturbations and persistent weak perturbations.
These results clearly show that  the response of a genetic switch to a
perturbation (or signal)  is a dynamic property which depends not only on the
(integrated or peak) pulse strength of the perturbation but also on its pulse
shape.

\section{Discussion }\label{sec:Discussion}

\subsection{General features}

The NS-FFS scheme has a number of characteristic features. First, like FFS,
NS-FFS does not perturb the given dynamical equations of the system (i.e. no
biasing force is applied). Instead, NS-FFS generates a 
biased ensemble of unbiased trajectories by proliferating those that move 
in a preferred direction and terminating those
that do not, with appropriate reweighting.
This means that NS-FFS, like FFS, is suitable for systems for which the
dynamical equations cannot easily be biased and reweighted  (e.g. because they
do not obey detailed balance). It also makes  NS-FFS highly suitable for
implementation as a wrapper around existing simulation code.

Second, NS-FFS generates trajectories with a complete history from $t=0$
on. This means that no
assumption of memory loss is made when a trajectory passes between interfaces,
as in some other rare event simulation methods  \cite{moroni04,faradjian04}.
Perhaps more significantly, NS-FFS also does not assume loss of memory on
reentry to the $A$ state. Most existing rare-event methods for stationary
systems, such as TPS, TIS, and FFS \cite{bolhuis02,erp03,allen06a}, rely on 
the assumption that trajectories which re-enter $A$ equilibrate rapidly
and can be treated as new, independent trajectories when they eventually
re-exit $A$. For most stationary systems where $\tau_A\ll\tau_{AB}$ 
this is a reasonable assumption, but for non-stationary
systems, such as those with non-Markovian or time-inhomogeneous macroscopic
switching dynamics \cite{becker12}, correlated entrance and re-exit from a
basin can make a significant contribution to the time-dependent quantity of
interest. Thus it is an essential feature of NS-FFS that memory loss is not
assumed  for $t>0$, even if the system re-enters the $A$ basin. 
\footnote{We note that if the microscopic dynamics of the
system has memory in itself, the history for $t<0$ may need to be specified as
part of the initial condition \cite{becker12}.}.

Third, once the NS-FFS algorithm has reached its steady state, each
interface bin emits one trajectory per started tree on average, so that
trajectories are sampled uniformly over the range $\roi$ (see
figs.~\ref{fig:barrierdensities},~\ref{fig:barrierabsdensities},
~\ref{fig:togg1}). 
Some fluctuations around this uniform sampling average
are tolerated in exchange for narrow weight distributions at each bin. These
features are a direct consequence of the basic flatPERM branching rule 
\cite{prellberg04}, which does not implement a negative feedback on 
(unweighted) trajectory numbers. NS-FFS is thus a `weak' flat-histogram method.

Fourth, we note that the effectiveness of a simulation scheme depends not only
on its ability to generate many samples but also on the independence of these
samples. Clearly, in NS-FFS, after a branching event, child trajectories remain
correlated for a certain time. This suggests that one should allow further
branching of the children only after a refractory time of the order of the
typical decorrelation time. We did not observe this modification to produce any
significant improvement for the systems studied here. This is presumably
because these systems are sufficiently stochastic that branched trajectories
anyway decorrelate rapidly between interfaces.  In contrast, it turned out
to be crucial to control the exponential growth of trajectory trees in the
initial phase of a simulation since these generate highly correlated
samples, which delay convergence of the crossing weight histogram. This
is simply and efficiently accomplished by using weight limits, as described
above.

\subsection{Progress coordinate vs.~time based branching}

The $\lambda$-if and $t$-if interface setups
(sec.~\ref{sub:Interfaces-in-phase-time}) differ in that the former equalizes
number fluxes of trajectories in $\lambda$-direction across the region $\roi$ 
while the latter equalizes their number density. Nevertheless, it
is of course possible to use either scheme to measure any quantity in a given
physical system; the choice of setup will affect only the efficiency of the
calculation.  We now briefly discuss how the observable of interest and the
computational overhead associated with branching can affect the  choice of the
most suitable setup.

\paragraph{Target observable}

Suppose one wishes to measure the time-dependent propensity $k_{AB}(t)$ for
exit over some final level $\lambda_{L}$ of the progress coordinate at the
boundary of state $B$, over some time interval $[t_{1},t_{I}]$
(cf.~Fig.~\ref{fig:timescales}). If re-crossings back from $B$ can be safely
neglected, we may place an absorbing boundary at $\lambda_{L}$. It is then
natural to use the $\lambda$-if setup, placing $\lambda$-interfaces at levels
$\{\lambda_{l}\}_{l=1\dots L}$, and count forward crossings only. In the limit
of slow escape over $\lambda_{L}$ such that the survival probability $\int
p(\lambda,t)\dd\lambda\simeq1$ over the time of observation, the observable
$k_{AB}$ coincides with the positive flux over the last interface $\lambda_{L}$
\cite{becker12}. Since the relative error in the positive flux is equalized
over all preceding interfaces, the $\lambda$-if setup will generate uniform
sampling for $k_{AB}(t)$ at the final interface over the time interval of
interest.

Alternatively, one may be interested in a potential of mean force $-\log
p(\lambda,t)$ over a region $\roi=[\lambda_{1},\lambda_{L}]\times[t_{1},t_{I}]$
in $\lambda-t$ space. In that case the $t$-if setup is the more natural choice,
since it generates a uniform relative error in $p(\lambda,t)$ over all bins.
One then obtains an estimate of $-\log p$ with uniform absolute error over the
region $\roi$. In particular, saddle points and basins are sampled with equal
frequency.

\paragraph{Branching overhead}

The $\lambda$-if and $t$-if setups differ in the relative overhead of
coordinate evaluation and branching. Clearly, the detection of crossings 
requires
the evaluation of $\lambda$, which incurs some computational overhead
$o_{\lambda}$ per evaluation. The branching move itself and the maintenance of
the tree structure add a second type of overhead $o_{b}$, which we also take to
be constant per branching event.

In the $t$-if setup, $\lambda$ evaluation and branching are coupled
and happen once every time-interface spacing $\tau_{t}$; the overhead
is $(o_{\lambda}+o_{b})/\tau_{t}$ per unit simulated time. In
contrast, in the $\lambda$-if setup, they are decoupled: interface
crossings are checked at intervals $\tau_\lambda$ while the trajectory
is being propagated, whereas trajectory branching happens only if a
crossing is detected.  The $\lambda$-evaluation interval
$\tau_{\lambda}$ is an independent adjustable parameter (but must not
be too long or the algorithm will fail to detect crossings). If
interface crossings happen every $\tau_{b}$ on average, the overhead
is $o_{\lambda}/\tau_{\lambda}+o_{b}/\tau_{b}$ per simulated time.

If $\lambda$ evaluation is expensive $(o_{\lambda}\gg o_{b})$, it is useful to
increase $\tau_{\lambda}$ or $\tau_{t}$, respectively, as far as 
possible without degrading the weight statistics. 
In the opposite case that $\lambda$ evaluation is cheap
($o_{\lambda}\ll o_{b}$), the $\lambda$-if setup may be advantageous, since
this setup gives the option to check
$\lambda$ often ($\tau_{\lambda}\ll\tau_{b}$), at small cost. 
Especially in systems
where  excursions towards the $B$ state tend to be short-lived, 
it is advantageous to check for interface crossings
significantly more often than they actually occur, since this increases 
the chance of detecting and capitalizing on short-lived forward excursions.
The dependence of NS-FFS performance on parameters will be
addressed more fully in a future publication.

\subsection{Variants and extensions of the algorithm}

\paragraph{Relation to stationary FFS}\label{sub:Stationary-version}

If NS-FFS is used to simulate a system which is in stationary state,  early and
late barrier crossings are equivalent; crossing statistics can then be improved
by binning early and late crossings together. This can be achieved within the
$\lambda$-if setup: trajectories are started from a stationary distribution at
$t=0$, and a single time bin is defined ($I=1$), leading to quicker convergence
of the, now one-dimensional, crossing weight histogram $H=\{H_{l1}\}_{1\leq
l\leq
  L}$.  This stationary version of NS-FFS is similar but not identical
to the `branched growth' variant of conventional FFS \cite{allen09}. In
branched-growth FFS, trees have a fixed number of children at each $\lambda$
interface crossing and are terminated exclusively at the next interface or when
the system returns to $A$; the stationary NS-FFS can be seen as an adaptive
generalization of this scheme.

\paragraph{Multiple progress coordinates}

In problems with multiple alternative  transition pathways, finding a single
good progress coordinate can be challenging. Use of a poor progress coordinate
fails to distinguish between  trajectories which are likely and unlikely to
result in a transition, making successful biasing of  transition paths
impossible. To some degree, NS-FFS already alleviates this problem compared to
TIS or FFS:
the extra dimension of time acts as a second progress coordinate which allows
us to discriminate between early and late crossings. In a class of systems
which show distinct `slow' and `fast' pathways of a reaction between $A$ and
$B$, an NS-FFS simulation would be able to separately enhance these pathways.

If different transition pathways share a common time scale then time as an
additional progress coordinate is not useful in itself, but one may be able to
find a small set of progress coordinates which successfully separate different
pathways. In this case, one can make a straightforward  generalization of the
NS-FFS scheme to multiple progress coordinates. As in any multi-dimensional
scheme, this will come at the cost of more bookkeeping and possibly slower
convergence of the weight histogram $H$ due to the increased number of bins,
see app.~\ref{sec:Multi-dimensional-progress-coord}.

\paragraph{Parallel version}

The NS-FFS algorithm lends itself to a parallel implementation. Each trajectory
can be simulated independently until an interface is crossed. At this point,
the shared histogram $H_{li}$ is read and updated. After branching, $n$
parallel simulations for the children are spawned. The communication between
simulation processes is restricted to histogram updates; depending on available
bandwidth these updates could also be cached and applied in groups, without
biasing the sampling. As the global histogram converges, updates become
unnecessary and the simulation gradually becomes trivially parallel.

\paragraph{Adaptive generalizations}

As shown in app.~\ref{sec:correctness}, branching/pruning events may be
introduced at will as long as weight is conserved on average
(Eq.~\ref{eq:unbiased}). This includes complete freedom of: adaptive updates of
the bin boundaries or the interface placement; inserting or removing
interfaces; smoothing of bin counts within or across interfaces; or pre-filling
of crossing histograms based on prior knowledge.

All of these options should allow for further performance improvement in
particular situations, to be explored in future work. In particular, it is
interesting to ask if an optimal interface and bin arrangement can be found
iteratively, as has been proposed for FFS \cite{borrero08}. 
A promising direction might be to monitor the local sampling noise and
adapt the interface arrangement in response to it.

\section{Conclusion}

In this article, we have introduced an enhanced sampling scheme, called NS-FFS,
which is conceptually simple allows the efficient sampling of rare events in
 non-Markovian and non-stationary systems.

The NS-FFS algorithm builds on two widely used ingredients: a flat-histogram
branched growth algorithm closely related to PERM
\cite{prellberg04,grassberger97}, and the concept of phase-space interfaces
\cite{erp03} to monitor progress towards a transition. NS-FFS is a
generalization of FFS \cite{allen06a}, and is straightforward to implement,
especially when one does not want to store the trajectories themselves.
We have demonstrated the correctness of the method, and given several simple
example applications which highlight both the effectiveness of the method and
the relevance of intrinsically time-dependent rare events.

A host of physical, chemical and biophysical problems are amenable to NS-FFS
simulations. These include the computation of time-dependent transition rates
in systems with time-dependent external driving \cite{becker12} such as the
signal-induced flipping of genetic switches studied here, crystal nucleation
during a temperature quench or protein unfolding under force; non-exponential
switching time distributions in processes that can be coarse-grained as
switches with memory, such as the switching of the bacterial flagellar motor
\cite{albada09,becker12}; and escape probabilities from a non-equilibrium
distribution in a metastable initial state within a prescribed `window of
opportunity', like the flipping of genetic switches induced by transient
pulses.

\subsection*{Acknowledgments}
We thank Daan Frenkel and Peter Bolhuis for many useful discussions.  This work
is part of the research program of the ``Stichting voor Fundamenteel Onderzoek
der Materie (FOM)'', which is financially supported by the ``Nederlandse
organisatie voor Wetenschappelijk Onderzoek (NWO)''. RJA is supported by a
Royal Society University Research Fellowship and by EPSRC under grant number
EP/I030298/1.

\appendix

\section{Reweighting criterion}\label{sec:correctness} 

In this section we  sketch  a proof that the condition of weight conservation
on average over branching outcomes, Eq.~\ref{eq:unbiased}, is sufficient for
unbiased sampling in a general branched-tree simulation scheme.
This amounts essentially to careful bookkeeping. No assumptions are made about
stationarity or loss of memory.

\subsection{Statement of the problem}

Consider a stochastic process $x(t)$ which is started at $t=0$ with a value of
$x(0)$ drawn from some initial distribution.
The process is fully characterized by all of its $m$-point joint probability
density functions (pdfs)
\begin{equation}
p(x_{m},t_{m};\dots;x_{1},t_{1})=
\E{\delta(x_{t_{m}}-x_{m})\cdots\delta(x_{t_{1}}-x_{1})},\label{eq:jointprob}
\end{equation}
for the trajectory to pass by the sequence of sample points
$x_{1},x_{2}\dots,x_{m}$
at times $0\leq t_{1}<t_{2}<\dots<t_{m}$. Here $x_{t_{m}}$ is a shorthand for
$x(t_m)$. In a brute-force simulation, the joint pdf is estimated by an average
over $S$ independent runs,
\begin{equation}
p(x_{m},t_{m};\dots;x_{1},t_{1})\simeq
  \frac{1}{S}\sum_{s=1}^{S}\delta(x_{t_{m}}^{s}-x_{m})
  \cdots\delta(x_{t_{1}}^{s}-x_{1}).\label{eq:jointbfsamp}
\end{equation}
(we note that the right hand side of this equation is singular; the approximate
equality is implied when Eq.~\ref{eq:jointbfsamp} is integrated over
finite regions).

We now introduce a single branching event at an intermediate time $t',$ and let
$m'=\max\{m|t_{m}<t'\}$. Upon branching, $n'$ statistically identical copies of
the system with independent futures are generated. That is, for a given history
up to time $t'$, each of the copies has the same conditional pdf
$p(x_{m},t_{m};\dots;x_{m'+1},t_{m'+1}|x_{m'},t_{m'};\dots;x_{1},t_{1})$ to
visit future points of phase space, but future points of different children are
mutually independent. No Markov assumption about the system is being made. The
task is now to assign correct weights to the child branches in order to
guarantee unbiased sampling.

\subsection{Unbiased sampling with one branching event}
\label{sub:Unbiased-sampling-with}

An obvious choice for the weights is to conserve the total in- and outgoing
weight at the branching point and to treat children equally. This rule gives a
weight $1/n'$ for each child trajectory from $t'$ on.

However, this rule disallows pruning, i.e.~$n'$=0. To enable pruning, it is
necessary to relax strict weight conservation at the branch point. To do this,
child numbers $n'=0,1,\dots n\ts{max}$ are drawn at random with probabilities
$b(n')$, $\sum_{n'=0}^{n\ts{max}}b(n')=1$. We then assign weights $w'_{c}$ to
the child branches $c=1\dots n'$ (if any). The \emph{expected} total weight of
trajectories passing through a sequence of points is then
\begin{align}
W(x_{m},t_{m};\dots;x_{1},t_{1})\nonumber \\
=\Bigl\langle\sum_{c=1}^{n'}w'_{c}\delta(x_{t_{m}}^{c}-x_{m})\cdots
  \delta(x_{t_{m'+1}}^{c}-x_{m'+1})\nonumber \\
\times\delta(x_{t_{m'}}-x_{m'})\cdots
  \delta(x_{t_{1}}-x_{1})\Bigr\rangle\nonumber \\
=\BE{\sum_{c=1}^{n'}w'_{c}}\E{\delta(x_{t_{m}}-x_{m})\cdots
  \delta(x_{t_{1}}-x_{1})}\nonumber \\
=\BE{\sum_{c=1}^{n'}w'_{c}}p(x_{m},t_{m};\dots;x_{1},t_{1}),\label{eq:totalwt}
\end{align}
where $m'$ is defined as above, sums running from $n'=1$ to $n'=0$ vanish by
definition, and we used the fact that children are identical. Clearly the
condition of weight conservation on average,
\begin{equation}
1=\BE{\sum_{c=1}^{n'}w'_{c}}=\sum_{n'=1}^{n\ts{max}}b(n')
  \sum_{c=1}^{n'}w'_{c},\label{eq:unbiasedgen}
\end{equation}
is necessary and sufficient for unbiased sampling, since then $W\equiv p$,
i.e.~the reweighting has corrected the bias.

Since there is no reason to treat child branches differently, we set
$w_{c}'=w'=r(n)w$ where $w=1$ is the parent weight. Eq.~\ref{eq:unbiasedgen}
now reduces to
\begin{equation}
1=\E{n'r(n')}=\sum_{n'=1}^{n\ts{max}}b(n')n'r(n'),\label{eq:unbiasedcond}
\end{equation}
Eq.~\ref{eq:unbiasedcond} constrains the choice weight factors $r$
for a given branching distribution $b$. For instance, if $n\ts{max}=2$,
and $(b(n))_{n=0,1,2}=(.5,.2,.3)$ then the choices $(r(n))_{n=1,2}=(0,5/3)$,
$(5,0)$ and $(5/4,5/4)$, are all unbiased.

Generalizing Eq.~\ref{eq:jointbfsamp}, we can then estimate $p$ from a branched
simulation of $S$ independent trees as
\begin{multline}
p(x_{m},t_{m};\dots;x_{1},t_{1})\\
\simeq\frac{1}{S}\sum_{s=1}^{S}r(n'_{s})
  \sum_{c_{s}=1}^{n'_{s}}\delta(x_{t_{m}}^{c_{s}}-x_{m})\cdots
  \delta(x_{t_{m'+1}}^{c_{s}}-x_{m'+1})\\
 \times\delta(x_{t_{m'}}^{s}-x_{m'})\cdots
  \delta(x_{t_{1}}^{s}-x_{1})\label{eq:treesamp}
\end{multline}
The first line on the rhs contains sample points after branching into $n'_{s}$
children (if any), and the second line those before branching (if any); the
weight factors $r$ satisfy eq\@.~\ref{eq:unbiasedcond}. Note that repeated
simulations average over not only the system but also the branching randomness.
For instance, the total weight of all trajectories at a given time is conserved
only on average over branching outcomes.

So far we have shown that a single stochastic branching move at time $t'$ and
re-weighted sampling according to Eq.~\ref{eq:treesamp} does not introduce a
bias if the child number probabilities $b(n')$ and child weight factors $r(n')$
obey the condition of weight conservation on average,
Eq.~\ref{eq:unbiasedcond}. We note that to verify this numerically, one would 
need to
generate many trees always with branching time $t'$ and count joint hits of all
bins around the points $x_{1},\dots,x_{m}$ at the times $t_{1},\dots,t_{m}$; if
the final time is larger than the branching time, hits are re-weighted the
factor $r(n_{s})$ appropriate for the respective branch number $n_{s}$.

\subsection{Adaptive branching probabilities}

We now show that the branching probabilities may be adapted according to an
arbitrary protocol. To see this, let $b$ depend on an arbitrary parameter
$\alpha$ such that $\sum_{n'}b_{\alpha}(n')=1$, and choose $r_{\alpha}$
accordingly, such that $\sum_{n'}b_{\alpha}(n')n'r_{\alpha}(n')=1$ for all
values of $\alpha$. Then, following the preceding discussion, a branching event
is unbiased for any $\alpha$. We may take $\alpha$ to be any random variable,
depending on a set of bias control parameters $\beta$ via some density
$\rho(\alpha|\beta)$. Since
Eq.~\ref{eq:totalwt} now reads
\begin{multline}
W(x_{m},t_{m};\dots;x_{1},t_{1})\\
=\int\dd\alpha\rho(\alpha|\beta)
\sum_{n'}b_{\alpha}(n')n'p(x_{m},t_{m};\dots;x_{1},t_{1})r_{\alpha}(n')\\
=\int\dd\alpha\rho(\alpha|\beta)p(x_{m},t_{m};\dots;x_{1},t_{1})\\
=p(x_{m},t_{m};\dots;x_{1},t_{1}),\label{eq:paramunbiased}
\end{multline}
unbiased sampling is still guaranteed. We thus have complete freedom to
introduce dependency of the branching probabilities on arbitrary extra
information, including the past or future system state, or the history of the
simulation.

\subsection{Multiple branching events}

Finally, we generalize the arguments above to multiple independent and
non-synchronized branching events on different branches of a tree. The idea is
that introducing a new branching event at a time $t''$ on an existing child
branch is bias-free, as long as the new $b$ and $r$ fulfill
Eq.~\ref{eq:unbiasedcond}. We then argue by induction over tree generations.

Each new branching event with $n^{l}$ children adds a weight factor $r(n^{l})$,
so that the instantaneous weight along a trajectory passing by the branching
events $(n',t';n^{2},t^{2};\dots;n^{k},t^{k})$ becomes
\begin{equation}
w(t;n',t';\dots;n^{k},t^{k})=\prod_{l:t^{l}<t}r(n^{l});\label{eq:multiweight}
\end{equation}

The system density estimate Eq.~\ref{eq:treesamp} now takes the form of a
hierarchical sum over all weighted branches existing at the final time. This is
simplest to write down using a recursive definition. We index a particular
branch (trajectory segment) anywhere in the tree by the sequence of the child
indices starting from the root, $\gamma=(c,c',\dots,c^{k})$, and let
$|\gamma|=k+1$ denote its nesting depth. The zeroth child index $c^{0}=c\equiv
s$ is the tree index. The points along the trajectory from the root up to and
including $\gamma$ are denoted $x_{t}^{\gamma}$ . Let
\begin{equation}
\pi^{\gamma}(x_{m},t_{m};\dots;x_{1},t_{1})=
  \prod_{\tilde{m}=1}^{m}\delta(x_{t_{\tilde{m}}}^{\gamma}-x_{\tilde{m}}),
  \label{eq:pimax}
\end{equation}
 if $\gamma$ has no children, and define recursively
\begin{equation}
\pi^{\gamma}(x_{m},t_{m};\dots;x_{1},t_{1})=r_{\gamma}
  \sum_{c=1}^{n_{\gamma}}\pi^{(\gamma,c)}(x_{m},t_{m};\dots;x_{1},t_{1}),
  \label{eq:pirec}
\end{equation}
if $\gamma$ has $n_{\gamma}$ children with weight factor $r_{\gamma}$. This
recursion terminates since the simulation has a finite maximal nesting depth
$K$. Note that $\pi^{\gamma}(x_{m},t_{m};\dots;x_{1},t_{1})$ counts the
weighted hits of $\gamma$ and all its descendants to the points
$(x_{m},t_{m};\dots;x_{1},t_{1})$. The weight is effectively given by
Eq.~\ref{eq:multiweight}.

All $S$ simulated trees together can then be represented as the descendants of
the empty path $()$. If we define $r_{()}=1/S$, $n_{()}=S$ and $|()|=0$, the
estimate for the joint $m$-point density is
\begin{equation}
p(x_{m},t_{m};\dots;x_{1},t_{1})\simeq
  \E{\pi^{()}(x_{m},t_{m};\dots;x_{1},t_{1})}.\label{eq:multitreesampling}
\end{equation}

Since Eq.~\ref{eq:pimax} coincides with the brute force estimate of the
density, and Eq.~\ref{eq:pirec} is unbiased by the arguments in
sec.~\ref{sub:Unbiased-sampling-with}, it follows by induction over $K$ that
the density estimate Eq.~\ref{eq:multitreesampling} is unbiased.

Eqs.~(\ref{eq:pimax},~\ref{eq:pirec},\ref{eq:multitreesampling}) directly
translate into a histogramming algorithm, which recursively parses a trajectory
tree while binning weighted counts. Binning can be carried out on-line.
This is straightforward if only the one-point density $p(x,t)$ is desired; 
online updates of $m$-point densities would require $m-1$ nested inner loops 
over the tree for each simulation step.

As a corollary, arbitrary $m$-point observables $A$ can be estimated along the
same lines. One just has to replace $\pi$ by $\pi_{A}$ in Eq.~\ref{eq:pirec}
and~\ref{eq:pimax} by
\begin{equation}
\pi_{A}^{\gamma}(t_{m};\dots;t_{m})=A(x_{t_{m}},\dots,x_{t_{1}}),
\label{eq:Amax}
\end{equation}
if $\gamma$ has no children. For instance, a two-point autocorrelation
function would correspond to 
$A(x_{t_{2}},x_{t_{1}})=x_{t_{2}}x_{t_{1}}-\E{x_{t_{2}}}\E{x_{t_{1}}}$.

\section{Variance of weighted bin counts}\label{sub:Variance-of-weighted}

Here, we write the variance of the total weight $W=\sum_{a=1}^{N}w_{a}$
accumulated in a given bin $B_{li}$ in terms of the statistics of the
trajectory weights $w_{a}$ and the number $N$ of trajectories which has reached
bin  $B_{li}$. We recall the `law of total variance':
\begin{equation}
\E{A^{2}}-\E A^{2}=\E{\E{A^{2}|\mathcal{C}}
  -\E{A|\mathcal{C}}^{2}}+(\E{\E{A|\mathcal{C}}^{2}}
  -\E{\E{A|\mathcal{C}}}^{2})\label{eq:total-variance}
\end{equation}
Here the inner expectation values are conditioned on some event $\mathcal{C}$,
and the outer expectations average over $\mathcal{C}$. We will denote the
conditional variance as 
$\E{\delta X^{2}|\mathcal{C}}\equiv\E{X^{2}|\mathcal{C}}-\E{X|\mathcal{C}}^{2}$
for any observable $X$. Eq.~\ref{eq:total-variance} then becomes $\E{\delta
A^{2}}=\E{\E{\delta A^{2}|\mathcal{C}}}+\E{\delta\E{A|\mathcal{C}}^{2}}.$

Consider an idealized NS-FFS simulation in which trajectories arriving at
$B_{li}$ are uncorrelated. Specifically, we assume that the incoming trajectory
weights $\{w_{a}\}$ are mutually independent and independent of $N$, and that
arrivals are a Poisson process. For the mean total crossing weight we then
obtain $\E W=\E N\E{w_{a}}.$ The total weight variance is given by
\begin{eqnarray*}
\E{\delta W^{2}} & = & \E{\E{\delta W^{2}|N}}+\E{\delta\E{W|N}^{2}}\\
 & = & \E N\E{\delta w_{a}^{2}}+\E{w_{a}}^{2}\E{\delta N^{2}}\\
 & = & \E N(\E{\delta w_{a}^{2}}+\E{w_{a}}^{2});
\end{eqnarray*}
the relative variance of the collected weight becomes 
\[
\frac{\E{\delta W^{2}}}{\E W^{2}}=\frac{1}{\E N}
  \left[1+\frac{\E{\delta w_{a}^{2}}}{\E{w_{a}}^{2}}\right].
\]

In the more general case, we now assume that correlations between branches
increase the noise while respecting the same scaling with the count number, and
write
\begin{equation}
\frac{\E{\delta W^{2}}}{\E W^{2}}\simeq
  \frac{\alpha_{N}}{\E N}\left[1+\alpha_{w}
  \frac{\E{\delta w_{a}^{2}}}{\E{w_{a}}^{2}}\right]
  \label{eq:weightvar}
\end{equation}
where $\alpha_{N}>1$ if trajectories arrive in bunches and $\alpha_{w}>1$
if their weights are correlated. The crossing flux $j_{li}$ is estimated as
$W_{li}/S$ and thus has the same noise, Eq.~\ref{eq:weightvar}, as $W_{li}.$

The noise in bin weights can be split up further. The incoming weights
$\{w_{a}\}$ are distributed with a mean and variance which result from both the
inter-bin variance between starting bins $B_{l'i'}$ and from the intra-bin
variances of outgoing weights from within $B_{l'i'}$. We have
\begin{eqnarray}
\E{w_{a}} & = & \E{\E{w_{a}|l'i'}}
  \text{, and using Eq.\,\ref{eq:total-variance},}\nonumber \\
\E{\delta w_{a}^{2}} & = & \E{\E{\delta w_{a}^{2}|l'i'}}
  +\E{\delta\E{w_{a}|l'i'}^{2}};\label{eq:totalvarwts}
\end{eqnarray}
here the first term is the mean intra-bin variance within originating bins, and
the second term is the inter-bin variance. Plugging in Eq.~\ref{eq:totalvarwts}
we can write the noise in $W$ as
\begin{equation}
\frac{\E{\delta W^{2}}}{\E W^{2}}=
  \frac{\alpha_{N}}{\E N}\left[1+\alpha_{w}
  \left\{ \frac{\E{\E{\delta w_{a}^{2}|l'i'}}}{\E{w_{a}}^{2}}
  +\frac{\E{\delta\E{w_{a}|l'i'}^{2}}}{\E{w_{a}}^{2}}\right\} \right].
  \label{eq:Wnoise}
\end{equation}

As the simulation progresses, crossing flux estimates converge, so that all
trajectories leaving $B_{l'i'}$ are eventually assigned the same weight. The
intra-bin term $\E{\E{\delta w_{a}^{2}|l'i'}}/\E{w_{a}}^{2}\propto\E{\delta
j_{l'i'}^{2}}/\E{j_{l'i'}}^{2}$ thus vanishes as $N\to\infty$. The inter-bin
term $\E{\delta\E{w_{a}|l'i'}^{2}}/\E{w_{a}}^{2}$ reflects the non-uniform
transition probabilities between bins and persists also in steady state.

In order to balance noise contributions, it seems reasonable to choose bin size
and interface spacing such that
$\alpha_{w}\E{\delta\E{w_{a}|l'i'}^{2}}\simeq\E{w_{a}}^{2}$ in an equilibrated
simulation.

\section{Multi-dimensional progress
coordinates}\label{sec:Multi-dimensional-progress-coord}

A multi-coordinate NS-FFS simulation can be set up as follows. First, find $K$
progress coordinates $\{\lambda^{k}\}_{k=1\dots K}$, with corresponding sets of
levels $\{\lambda_{0}^{k}<\dots<\lambda_{L_{k}}^{k}\}_{k=1\dots K}$. Denote the
interval $(\lambda_{l-1}^{k},\lambda_{l}^{k})=\Lambda_{l}^{k}$. Define a subset
of $K'\leq K$ progress coordinates $\{\lambda^{k'}\}_{k'=1\dots K'}$. Only the
first $K'$ sets of interfaces will trigger branching events; for these
interfaces, define bins:
\begin{multline}
B_{l_{1}l_{2}\dots l_{k'}\dots l_{K}}^{k'}=\\
\{(x,t)|\lambda^{k'}(x,t)=\lambda_{l_{k'}}^{k'}
\text{ and }\lambda^{k}(x,t)\in\Lambda_{l_{k}}^{k}\text{ for }k\neq k'\}
\label{eq:Bmulti}
\end{multline}
Then, proceed as before: branching moves are triggered on the first $K'$
interfaces, based on the corresponding crossing weights $H_{l_{1}l_{2}\dots
l_{k'}\dots l_{K}}^{k'}$.

In this setting, time is treated as another progress coordinate. To recover the
NS-FFS setups discussed in sec.~\ref{sec:Nonstationary-Flux-Sampling} in this
setting, let $K=2,K'=1$. For the $\lambda$-if setup, set $\lambda^{1}=\lambda$,
$\lambda^{2}=t$, $L_{1}=L$ and $L_{2}=I$; for the $t$-if setup, set
$\lambda^{1}=t$, $\lambda^{2}=\lambda$, $L_{1}=I$ and $L_{2}=L$.

\section{Piecewise linear potential}\label{sec:Piecewise-linear-potential}

We solve the Fokker Planck equation associated with Eq.~\ref{eq:langevin}
in Laplace space,
\begin{equation}
sp-p_{0}+\partial_{x}(\mathrm{sgn}(x)aDp-D\partial_{x}p)=0,\label{eq:fplap}
\end{equation}
where $p=p(x,s)=\int_{0_{-}}^{\infty}p(x,t)e^{-st}\dd t$ is the Laplace
transformed density, and the initial condition $p_{0}(x)=p(x,t=0_{-})\equiv0$.
Particles are injected at $t=0$; the boundary conditions for the
total flux $j(x,t)=\mathrm{sgn}(x)aDp(x,t)-D\partial_{x}p(x,t)$ at
the left boundary read 
\begin{align}
j(x=-1,t)=\delta(t) & \text{ or in Laplace space},\nonumber \\
j(x=-1,s)=1.\label{eq:bconds}
\end{align}
They incorporate both the reflecting boundary and the injection of
unit probability at $x=-1$ at $t=0.$ 

At the right boundary, we consider either reflecting (referred to
as r/r) or absorbing (r/a) conditions: 
\begin{align}
j(x=1,s)=0, & \text{ or}\label{eq:bcondref}\\
p(x=1,s)=0 & \text{, respectively.}\label{eq:bcondsabs}
\end{align}

Eqs.~\ref{eq:fplap},~\ref{eq:bconds} and \ref{eq:bcondref}/\ref{eq:bcondsabs}
can be solved by using the ansatz $p(x,s)=e^{\frac{1}{2}ax}\tilde{p}(x,s)$,
and joining solutions in the regions $x<0$ and $x>0$. After straightforward
but lengthy algebra, the solution in the r/r case can be written as
\begin{multline}
p(x,s)=e^{-\frac{1}{2}a(1-|r|)}\times\\
\tfrac{qa\sinh(q(1-|r|))-2q^{2}\cosh(q(1-r))
  -\theta(-r)a^{2}\sinh(qr)\sinh(q)}{2sa\sinh^{2}(q)-4sq\sinh(q)\cosh(q)}
  \label{eq:ex lap ref}
\end{multline}
where $q=\sqrt{s/D+a^{2}/4}$ and $\theta$ is the unit step function. For the
r/a case we obtain
\begin{multline}
p(x,s)=e^{-\frac{1}{2}a(1-|r|)}\times\\
\tfrac{qa\sinh(q(1-r))-\theta(-r)
  \frac{a^{2}}{2}(\cosh(q(1+r)-\cosh(q(1-r)))}{2sa\sinh^{2}(q)+aDq^{2}}.
  \label{eq:ex lap abs}
\end{multline}
Both Green's functions have poles only on the non-positive real $s$-axis. The
barrier crossing time can be extracted by solving for the largest negative pole
at $-s\ts{AB}$; in the r/r case, $s\ts{AB}=k\ts{AB}+k\ts{BA}=2k\ts{AB}$ while
in the r/a case, $s\ts{AB}=k\ts{AB}$. If the barrier is high, we may expand the
relevant denominators in eqs.~\ref{eq:ex lap ref}, \ref{eq:ex lap abs} for
$\frac{|s|}{Da^{2}}\ll1$. We find in both cases (r/r and r/a) that the waiting
time scales exponentially with the barrier:
\begin{equation}
\tau\ts{AB}=s\ts{AB}^{-1}=\frac{2e^{a}}{a^{2}D}+O(1/a).\label{eq:barriertauw}
\end{equation}
In the r/a case we can also evaluate the exit flux through the absorbing
boundary,
\begin{equation}
j(x=1,s)=j\ts{AB}(s)=\frac{4q^{2}}{a^{2}-(a^{2}-4q^{2})\cosh(2q)}
\label{eq:exitflx}
\end{equation}

The equilibration time within basin $A$ can be estimated as a diffusion time
for covering the thermally accessible range of $x$; alternatively, a more
accurate pre-factor can be obtained by solving Eq.~\ref{eq:fplap} with
Eq.~\ref{eq:bconds} as above but replacing the W-shaped potential by a
uniformly increasing ramp potential $U=ax$ of the same slope. Evaluating the
slowest relaxation time now gives the relaxation time in $A$,
\begin{equation}
\tau_{A}=\frac{4}{a^{2}D}.\label{eq:barriertaua}
\end{equation}

Finally, the crossing time $\tau_{C}$ is the mean first passage time for
diffusion through region $C$; that is, from the boundary of region $A$ at
$-x_{C}$, up the barrier and down on the other side until reaching the boundary
of region $B$ at $x_{C}$, without returning to $A$. If we disregard
trajectories that cross $x=0$ more than once, $\tau_{C}$ is the sum of the mean
first passage times for the two segments, with negative and positive constant
drift, respectively. The non-intuitive but well-known result is that diffusion
of successful transition paths up the barrier takes as long as down the barrier
(see e.g.~\cite{redner01}). In the drift-dominated regime, the mean first
passage time is controlled by the drift velocity $aD$. One obtains
\begin{equation}
\tau_{C}=2\times\frac{x_{C}}{aD}\lesssim\frac{2}{aD}.\label{eq:barriertauc}
\end{equation}

\bibliographystyle{unsrt}
%\bibliography{../../NilsRef}

\end{document}